\documentclass[aps,pre,amsmath,amsfonts,amssymb,nofootinbib]{revtex4}

\usepackage{graphicx}
\usepackage{epsfig}

\begin{document}

\date{\today}

\title{On the number of circuits in random graphs}
\author{Enzo Marinari$^{\,1}$ and Guilhem Semerjian$^{\,2}$}
\affiliation{$^{1\,}$
Dipartimento di Fisica and INFN, Universit\`a di Roma La Sapienza, 
P. A. Moro 2, 00185 Roma, Italy,
}
\affiliation{$^{2\,}$
Dipartimento di Fisica and CNR, Universit\`a di Roma La Sapienza, 
P. A. Moro 2, 00185 Roma, Italy.
}

\begin{abstract} 
We apply in this article (non rigorous) statistical mechanics methods to 
the problem of counting long circuits in graphs. The outcomes of this approach 
have two complementary flavours. On the algorithmic side, we propose
an approximate counting procedure, valid in principle for a large class of 
graphs.
On a more theoretical side, we study the typical number of 
long circuits in random graph ensembles, reproducing rigorously known results
and stating new conjectures. 
\end{abstract}

\maketitle

\section{Introduction}

Random graphs~\cite{Bollo_book,Janson_book} 
appeared in the mathematical literature as a convenient
tool to prove the existence of graphs with a certain property: instead
of a direct constructive proof exhibiting such a graph,
one can construct a random ensemble of graphs
and show that this property is true with a positive probability.
Soon afterwards the study of random graphs acquired interest on
its own and led to many beautiful mathematical results. A large class
of problems in this field can be formulated in the following generic way: 
a graph $H$ being given,
what is the probability that a graph $G$ extracted from the random ensemble 
under consideration contains $H$ as a subgraph? 
With a more quantitative ambition,
one can define ${\cal N}_H(G)$ as the random variable counting the number of
occurrences of distinct copies of $H$ in $G$, and study its distribution.
These problems are relatively simple when the pattern $H$ 
remains of a finite size in the thermodynamic limit, i.e. when the size
of the random graph $G$ diverges. The situation can become much more involved 
when $H$ and $G$ have large sizes of the same order, as ${\cal N}_H(G)$ can
grow exponentially with the system size.

In this article we shall consider these questions when the looked for subgraph
$H$ is a long circuit (also called loop or cycle), i.e. a closed self-avoiding 
path visiting a finite fraction of the vertices of the graph. The level of
accuracy of the rigorous results on this problem depends strongly on the
random graph ensemble~\cite{Bollo_book,Janson_book,Wormald_review}. 
The regular case (when all vertices of the graph have
the same degree $c$) is the best understood one. It has for instance been 
shown that $c$-regular random graphs with $c \ge 3$ have with high probability 
Hamiltonian circuits (circuits which visit all vertices of the graph) and the
distribution of their numbers is known~\cite{Robinson_ham,Janson_ham}. 
This study has been generalized to circuits of all length in~\cite{Garmo}.
Less is known for the classical Erd\"os-R\'enyi ensembles, where the degree
distribution of the vertices converges to a Poisson law of mean $c$. Most 
results concerns either the neighborhood of the percolation transition at
$c=1$~\cite{Janson_distrib,Luczak,Flaj}, or the opposite limit of very 
large mean connectivity, either finite with respect to the size of the 
graph~\cite{Frieze} or diverging like its logarithm~\cite{Posa} (it is
in this latter regime that the graphs become Hamiltonian).
We shall repeatedly come back in the following on this discrepancy between
regular random graphs where probabilistic methods have been proved so 
successful and the other ensembles for which they do not seem powerful enough
and might be profitably complemented by approaches inspired by statistical
mechanics. We will discuss in particular a conjecture formulated by 
Wormald~\cite{Wormald_review}, according to which random graph ensembles with
a minimal degree of 3 (and bounded maximal degree) should be Hamiltonian with
high probability.

Besides this probabilistic point of view (what are the characteristics of
the random variable associated to the number of circuits), the problem
has also an algorithmic side: how to count the number of circuits in a given
graph? Exhaustive enumeration, even using smart algorithms~\cite{Johnson},
is restricted to small graphs as the number of circuits grows exponentially
with the size. More formally, the decision problem of
knowing if a graph is Hamiltonian (i.e. that it contains a circuit
visiting all vertices) is NP-complete~\cite{GaJo}. A probabilistic
algorithm for the approximate counting of Hamiltonian cycles is 
known~\cite{proba_Hamilton}, but is restricted to graphs with large 
minimal connectivity.

Random graphs have also been largely considered in the physics literature, 
mainly
in the real-world networks perspective~\cite{AlBa}, i.e. in order to compare 
the characteristics of observed networks, of the Internet for instance, with
those of proposed random models. Empirical measures for short loops in real 
world graphs were for instance presented in~\cite{BiCaCo}. Long 
circuits visiting a finite fraction of the vertices were also studied 
in~\cite{ben}. The behavior of cycles in the neighborhood of the percolation
transition was considered in~\cite{BenKra}, and the average number of
circuits for arbitrary connectivity distribution was computed in~\cite{BiMa}.

In this paper we shall turn the counting problem into a statistical 
mechanics model,
which we treat within the Bethe approximation. This will led us
to an approximate counting algorithm, cf. Sec.~\ref{sec_bethe}. We will
then concentrate on random graph ensembles and compute the typical number
of circuits with the cavity replica-symmetric method~\cite{MePa_Bethe}
in Sec.~\ref{sec_typical}.
The next two sections will be devoted to the study of the limits of
short and longest circuits, then we shall investigate the validity of the
replica-symmetry assumption in Sec.~\ref{sec_stability}.
We perform a comparison with exhaustive enumerations on small graphs in
Sec.~\ref{sec_enumerations} and draw our conclusions in 
Sec.~\ref{sec_conclusion}. Three appendices collect more technical 
computations. A short account of our results has been published in~\cite{EPL}.

\section{A statistical mechanics model and its Bethe approximation}
\label{sec_bethe}
\subsection{Derivation of the BP equations}

Let us consider a graph $G$ on $N$ vertices (also called sites in the 
following) $i=1,\dots,N$, with $M$ edges (or links) $l=1,\dots,M$. 
The notation $l=\langle i j \rangle$ shall mean that the
edge $l$ joins the vertices $i$ and $j$.
The degree, or connectivity, of a site is the number of links it belongs to.
The graphs are assumed in the main part of the text to be simple, i.e.
without edges from one vertex to itself or multiple edges between two vertices.
We denote $\partial i$ the set of neighbors of the vertex $i$, and use the
symbol $\setminus$ to subtract an element of a set: if $j$ is a neighbor of
$i$, $\partial i \setminus j$ will be the set of all neighbors of $i$
distinct of $j$. The same symbol $\partial i$ will be used for the set of
edges incident to the vertex $i$, the context will always clarify which
of the two meanings is understood. 

A circuit of length $L$ is an ordered set of $L$ different vertices, 
$(i_1,\dots,i_L)$, such that $\langle i_n i_{n+1}\rangle$ is an edge of the 
graph for all $n\in [1,L-1]$, as well as $\langle i_L i_1\rangle$. 
Two circuits are distinct if they do
not share the same set of edges (i.e. the starting point and the orientation
of a tour along the vertices is not relevant), and we denote
${\cal N}_L(G)$ the number of distinct circuits of length $L$ in a graph $G$.

The degrees of freedom of our model are $M$ variables $S_l\in\{0,1\}$ placed
on the edges of the graph, with their global configuration called
$\underline{S}=\{S_1,\dots,S_M\}$. We shall also use
$\underline{S}_i=\{S_l|l \in \partial i\}$ for the configuration
of the variables on the links around the vertex $i$. We introduce the
following probability law on the space of configurations:
\begin{equation}
p(\underline{S}) = \frac{1}{Z(u)} w(\underline{S}) \ , \qquad
w(\underline{S})=
\left( \prod_{l=1}^M \widehat{w}_l(S_l) \right)
\left( \prod_{i=1}^N w_i(\underline{S}_i) \right) \ ,
\label{eq_law}
\end{equation}
where $Z(u)$ is the normalization constant, and the weights $\widehat{w}_l$,
$w_i$ are given by
\begin{equation}
\widehat{w}_l(S_l) = u^{S_l} \ , \qquad
w_i(\underline{S}_i) = \begin{cases} 
1 \quad \mbox{if} \quad \underset{l \in \partial i}{\sum} S_l \in \{0,2\} \\
0 \quad \mbox{otherwise}
\label{eq_wi}
\end{cases} \ .
\end{equation}
By convention $w_i=1$ if $\partial i =\emptyset$, that is to say if the
vertex $i$ is isolated. 
The relevance of this model for the counting of circuits is unveiled by the 
following reasoning. Each configuration 
$\underline{S}$ can be associated to a subgraph of $G$, retaining
only the edges $l$ such that $S_l=1$. The probability (with respect to the law
(\ref{eq_law})) of such a subgraph is 
non zero only if the retained edges form closed circuits (any site $i$ is 
constrained by $w_i$ to be surrounded by either 0 or 2 edges of the subgraph), 
and in that case it is proportional to $u^L$ with $L$ the number of its
edges. This implies that the normalization factor $Z(u)$ is the
generating function of the numbers ${\cal N}_L(G)$,
\begin{equation}
Z(u) = \sum_L u^L {\cal N}_L(G) \ .
\label{eq_sumZ}
\end{equation}
A precision should be made at this point: we defined above a circuit as
a self-avoiding closed path. From the weights on the
configurations defined by Eqs.~(\ref{eq_law},\ref{eq_wi}), ${\cal N}_L(G)$ 
counts in fact the number of configurations made of possibly several 
vertex disjoint circuits, of total length $L$. In the following we shall
concentrate on the limit of large graph and of long circuits, and we 
expect the leading order behaviour of ${\cal N}_L$ not to be affected by this 
subtlety (see App.~\ref{app_disjoint} for a combinatorial argument in favour
of this thesis), that will be kept understood from now on\footnote{The 
reader might think this problem would be solved by enlarging the space of 
the configurations $S_l$ to a Potts-like spin,
$S_l\in \{0,1,\dots,q\}$, with the weight $w_i$ enforcing that either all
variables around $i$ are vanishing, or two are non zero and of the same 
colour $1,\dots,q$. In the bivariate generating function $Z(u,q)$ 
$q$ is then conjugated to the number of disconnected circuits, and
the limit $q \to 0$ should allow to eliminate configurations made of several
disconnected circuits. However, the Bethe approximation of this model
is pathological, and we shall not pursue this road here.}.
Note also that~\cite{MC_enumeration} proposed a Monte Carlo Markov Chain
algorithm for the evaluation of such a partition function.

We are thus performing a canonical computation where the length
of the circuits is allowed to fluctuate around a mean value fixed by
the conjugate external parameter $u$. In the
thermodynamic limit $N \to \infty$ the saddle-point method can be used to
evaluate the sum (\ref{eq_sumZ}). Defining  
$f(u)=\frac{1}{N} \ln Z(u)$ and
$\sigma(\ell) = \frac{1}{N} \ln {\cal N}_{L=\ell N}$, where $\ell = L/N$ is
a reduced intensive length, one obtains:
\begin{equation}
f(u) = \max_\ell [ \ell \ln u + \sigma(\ell) ] \ .
\label{eq_Legendre}
\end{equation}
In this limit the fluctuations of the intensive circuit length in the
canonical ensemble vanishes, $\ell$ is concentrated around its mean
value $\ell(u)= u f'(u)$.
The (concave hull of the) microcanonical entropy 
can thus be obtained from the canonical free-energy (with a slight abuse of
terminology we shall use this denomination for $f(u)$) by an inverse Legendre 
transform,
\begin{equation}
\sigma(\ell) = \min_u [ f(u) - \ell \ln u ] \ .
\label{eq_Legendre2}
\end{equation}

We shall now use the Bethe approximation to obtain an estimation of the
generating function $Z(u)$. We sketch first the general strategy to
derive Bethe approximations of statistical models (see~\cite{Yedidia} for a
comprehensive discussion), before applying it to the present case.

Consider a (non-negative) weight function $w(\underline{S})$ defined 
on a space of configurations $\{\underline{S}\}$. The computation of the
partition function, $Z= \sum_{\underline{S}} w(\underline{S})$, can be 
reformulated as an extremization problem. Indeed, the Gibbs functional 
free-energy,
\begin{equation}
F_{\rm Gibbs}[p_{\rm v}] = \sum_{\underline{S} } p_{\rm v}(\underline{S} ) 
\ln \left( \frac{p_{\rm v}(\underline{S} )}{w(\underline{S} )} \right) \ ,
\end{equation}
is minimal (in the space of normalized variational distributions)
for $p_{\rm v}(\underline{S})=p(\underline{S})=w(\underline{S})/Z$, where it 
takes the value $(-\ln Z)$.
In general finding the minimum of this functional is not simpler than a
direct computation of $Z$, however this formulation opens the way to 
variational approaches: the minimum of $F_{\rm Gibbs}$ in a
restricted set of trial distributions $p_{\rm v}$, more easily parametrized 
than generic ones, yields an upper bound on $(-\ln Z)$. The
simplest implementation of this idea is the mean-field approximation, in
which the trial distributions are factorized, 
$p_{\rm v}(\underline{S}) = \prod_l p_l(S_l)$. A natural refinement consists in
introducing correlations between neighboring variables in the trial 
distributions. Consider for instance a weight function of the form 
(\ref{eq_law}), for arbitrary $\widehat{w}_l$ and $w_i$. One can easily show
that if the underlying graph $G$ were a tree, the true probability distribution
would be given by
\begin{equation}
p(\underline{S}) = \left(\prod_{l=1}^M p_l(S_l) \right)^{-1}
\left(\prod_{i=1}^N p_i(\underline{S}_i) \right) \ ,
\end{equation}
where $p_l$ and $p_i$ are the exact marginals (for instance,
$p_l(S_l)=\sum_{\underline{S} \setminus S_l}p(\underline{S})$)
of the law $p$. 
When the
graph is not a tree, this expression is not valid any more. The Bethe
approximation consists however in assuming that trial probability 
distributions can be approximately written under this form even if the graph
contains cycles. This yields the so-called Bethe free-energy,
\begin{equation}
F_{\rm Bethe}[\{p_i\},\{p_l\}] = \sum_{i=1}^N \sum_{\underline{S}_i} 
p_i(\underline{S}_i) \ln \left( 
\frac{p_i(\underline{S}_i)}{w_i(\underline{S}_i)} \right) - 
\sum_{l=1}^M \sum_{S_l} p_l(S_l) \ln \left( p_l(S_l) \widehat{w}_l(S_l)\right)
\ .
\end{equation}
This free-energy is to be minimized with respects to the approximate marginals
$p_l$, $p_i$, which have to respect two types of constraints:
\begin{itemize}
\item[$\bullet$] $p_l$ and $p_i$ are normalized.
\item[$\bullet$] they are consistent, i.e. for each link 
$l=\langle i j\rangle$, one has
\begin{equation}
p_l(S_l) = \sum_{\underline{S}_i \setminus S_l} p_i(\underline{S}_i) =
\sum_{\underline{S}_j \setminus S_l} p_j(\underline{S}_j) \ .
\label{eq_consistency}
\end{equation}
\end{itemize}
This constrained minimization can be 
performed considering the $\{p_i\},\{p_l\}$ as independent, at the price of 
the introduction of Lagrange multipliers to enforce the conditions 
(\ref{eq_consistency}). 
It is well known that such a procedure amounts to look for a fixed point of 
the corresponding belief propagation (BP) equations~\cite{Yedidia}. 
In this setting the Lagrange multipliers are interpreted as messages sent 
by variables to neighboring constraints, and vice-versa.

Let us now apply this formalism to the specific weights defined
in Eq.~(\ref{eq_wi}). A peculiarity of $w_i$ has to be
kept in mind: it can be strictly vanishing when the geometrical constraint
of having 0 or 2 present edges around each vertex is not fulfilled. As a
consequence, the approximate variational marginals $p_i(\underline{S}_i)$
have to respect this constraint, and vanish when $w_i(\underline{S}_i)=0$.
The Bethe free-energy reads now
\begin{equation}
F_{\rm Bethe}[\{p_i\},\{p_l\}] = \sum_{i=1}^N \sum_{\underline{S}_i} 
p_i(\underline{S}_i) \ln \left( 
p_i(\underline{S}_i) \right) - 
\sum_{l=1}^M \sum_{S_l} p_l(S_l) \ln \left( p_l(S_l) u^{S_l}\right)
 \ ,
\label{eq_F_Bethe}
\end{equation}
where the convention $0 \ln 0 =0$ has been used,
for the strictly forbidden configurations with $w_i(\underline{S}_i)=0$ not to
contribute to $F_{\rm Bethe}$. 

\begin{center}
\begin{figure}
\includegraphics[width=4.2cm]{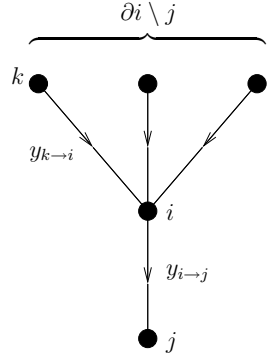}
\caption{The messages involved in Eq.~(\ref{eq_msgs}).}
\label{fig_eq_msg}
\end{figure}
\end{center}

A possible parametrization of the marginals achieving the extremum of the
Bethe free-energy is
\begin{equation}
p_l(S_l) = \frac{1}{C_l}(u \, y_{i \to j}\, y_{j \to i})^{S_l} \ , \qquad
p_i(\underline{S}_i) = \frac{1}{C_i} w_i(\underline{S}_i) 
\prod_{j\in \partial i} (u \, y_{j \to i})^{S_{\langle ij\rangle}} \ ,
\end{equation}
where the $C_l$ and $C_i$ are normalization constants, and for 
each link $\langle i j\rangle$ of the graph a pair of directed (real positive) 
messages has been introduced, $y_{i \to j}$ and $y_{j \to i}$. These
messages obey the following BP equations,
\begin{equation}
y_{i \to j} = \frac{u \underset{k\in \partial i \setminus j}{\sum} 
y_{k \to i}}{1+\frac{1}{2} u^2 
\underset{\underset{k \ne k'}{k,k' \in \partial i \setminus j}}{\sum} 
y_{k \to i} \ y_{k' \to i}} \ ,
\label{eq_msgs}
\end{equation}
cf. Fig.~\ref{fig_eq_msg} for a graphical representation. Roughly speaking, 
$y_{i \to j}$ is proportional to the probability that the edge 
$\langle i j\rangle$ would be present if the constraint $w_j$ and the
weight $u^{S_l}$ were to be discarded. Hence the form of 
Eq.~(\ref{eq_msgs}): the numerator corresponds
to the situation where $\langle i j\rangle$ is present, the constraint $w_i$
imposes then that exactly one of the edges of $\partial i\setminus j$ is
also present. The denominator states on the contrary that if 
$\langle i j\rangle$ is absent, either none or two of the edges of 
$\partial i\setminus j$ are present.
The normalization constants of the marginals are easily computed,
\begin{equation}
C_l = 1 + u \, y_{i \to j} \, y_{j \to i} \ , \qquad
C_i = 1+ \frac{1}{2} u^2 
\underset{\underset{k \ne k'}{k,k' \in \partial i}}{\sum} y_{k \to i} \
y_{k' \to i} \ ,
\end{equation}
and one can check, using the BP equations, that the consistency conditions
are indeed respected by these expressions of the marginals. 
Moreover the value of 
$F_{\rm Bethe}$ at its minimum can be expressed in terms of
the normalization constants $C_i$ and $C_l$. Using this value as an 
approximation for $-\ln Z(u)$ the free energy in the Bethe approximation
can be written as:
\begin{equation}
N f(u) = - \sum_{l=1}^M \ln(C_l) + \sum_{i=1}^N \ln(C_i) \ .
\label{eq_f}
\end{equation}
One should also compute the length of the circuits in the
configurations selected at a given value of $u$, 
$\ell(u)=u f'(u)$. It is rather unwise to use Eq.~(\ref{eq_f})
to compute the derivative $f'(u)$, as this expression involves the
messages which are solution of the BP equations 
and hence have a non trivial dependence
on $u$. On the contrary the expression (\ref{eq_F_Bethe}) being
variational, it is enough to compute its explicit derivative with respects
to $u$ to obtain:
\begin{equation}
\ell(u) = \frac{1}{N} \sum_{l=1}^M p_l(1) = \frac{1}{N} 
\sum_{\langle ij \rangle}
\frac{u \, y_{i \to j} \, y_{j \to i}}
{1 +u \, y_{i \to j} \, y_{j \to i}} \ .
\label{eq_ell}
\end{equation}
The first equality is natural, the average length of the circuits being
equal to the sum of the probabilities of presence of all the edges of
the graph. Note also that the marginal probabilities contain individually
some local information: for instance $p_l(1)$ is the fraction of circuits 
of length $\ell(u)$ which go through the particular link $l$.

\begin{center}
\begin{figure}
\includegraphics[width=13cm]{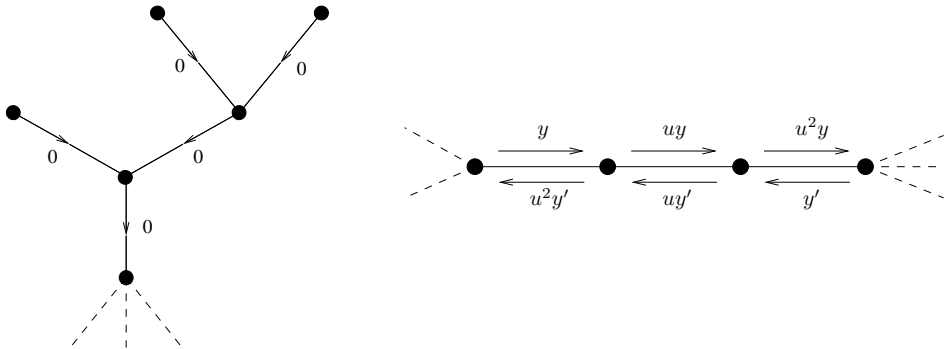} 
\caption{Left: the leaf removal procedure interpreted in terms of null 
messages. Right: the behaviour of messages along chains of degree 2 vertices.}
\label{fig_leaf_and_chain}
\end{figure}
\end{center}

Let us now come back for an instant on the BP equations and underline two
simple properties they possess. In Eq.~(\ref{eq_msgs}) we
used the natural convention that sums on empty sets are null. The first
consequence is that $y_{i \to j}=0$ if $j$ is the only neighbor of $i$, 
as $\partial i \setminus j = \emptyset$. In such a situation $i$ is indeed 
a leaf of the graph,
and no circuits can go through the edge $\langle ij\rangle$.
Even if in general the directed message in the reverse direction $y_{j \to i}$
is non zero, one can easily check that the edge $\langle ij\rangle$ does
not contribute to the free energy. In other words the physical observables
are unaffected by the leaf removal process, in which the graph $G$ is
deprived of the dangling edge $\langle ij\rangle$. Moreover this simplification
can be iteratively repeated, until no leaves are present in the remaining 
graph. An illustration of this process in terms of the null directed
messages is given in the left part of Fig.~\ref{fig_leaf_and_chain}. 
This property of the BP equations reflects
the fact that the circuits of a graph
are necessarily part of its 2-core, that is to say the largest of its subgraphs
in which all sites have connectivity at least 2.

Consider now a site $i$ with two neighbors $j$ and $k$, for which the BP
equations read $y_{i \to j} = u y_{k \to i}$ and 
$y_{i \to k} = u y_{j \to i}$. This implies that along a chain of
degree 2 sites, the directed messages follow a geometric progression, cf.
the right part of Fig.~\ref{fig_leaf_and_chain}, 
and in consequence one easily shows that the marginal 
probability of all edges in a chain are equal: if a circuit goes
through one of the edges of the chain, it must go through all of it.

\subsection{An approximate counting algorithm} 

The presentation of the Bethe approximation in terms of messages~\cite{Yedidia}
we followed in the previous section suggests in a very natural way
the following algorithm for the approximate counting of long circuits in a 
given graph:
\begin{itemize}
\item[$\bullet$]
initialize messages $y_{i \to j}$ for each directed edge of the graph to
some random positive values.
\item[$\bullet$] 
iterate the BP equations (\ref{eq_msgs}) at a given value of $u$ until 
convergence has been reached.
\item[$\bullet$]
using the messages solution of the BP equations,
compute $f(u),\ell(u)$ from Eqs.~(\ref{eq_f},\ref{eq_ell}), 
and $\sigma(\ell(u))=f(u) - \ell(u) \ln u $ (cf. Eq.~(\ref{eq_Legendre2})).
\item[$\bullet$]
repeat this procedure for different values of $u$ to obtain a plot of 
$\sigma(\ell)$ parametrized by $u$.
\end{itemize}

This algorithm is of course far from being exact. A first limitation is that
the BP equations are not a priori convergent, on the contrary it is 
easy to construct counter-examples of small graphs on which they do not 
reach any fixed point. 
It would thus be interesting to determine under which conditions 
the convergence towards an unique (non-trivial) fixed point is ensured. 
This kind of question has been the subject of recent interest, see for 
instance~\cite{Tati,Heskes}. Another possible criticism is that even in the 
case of convergence of the BP equations, the prediction for the number of 
loops relies on the Bethe approximation, which is an uncontrolled one. 
This being said, one should however keep in mind that for large
graphs with numerous circuits, an exact enumeration~\cite{Johnson} 
is computationally very expensive and reaches
very soon the limitations of present time computers. The approximate algorithm
we introduced here can then serve as an efficient alternative, even if its 
predictions should be treated with caution.

We presented in~\cite{EPL} the results of such a procedure when applied to
a real-world network of the Autonomous System Level description of the
Internet~\cite{DIMES}, 
allowing to estimate the total number of circuits, the length of
the most numerous circuits and the maximal length circuits, obtaining numbers
which are far beyond the possibilities of exhaustive counting. We also checked
the compatibility of our results with the direct enumeration of very short 
circuits.

\section{The typical number of circuits in random graphs ensembles}
\label{sec_typical}

\subsection{Definitions}

The rest of the paper shall be devoted to the study of the number of 
circuits in graphs $G$ belonging to random ensembles. In the regime we are
interested in (long circuits of large graphs with finite mean degree), the
common wisdom about the statistical mechanics of disordered systems is that
the random variable $\log({\cal N}_{L=N \ell})/N$ should be concentrated 
around its average, the quenched entropy $\sigma_{\rm q}(\ell)$. More
formally, one expects the existence of a constant 
$\ell_{\rm max}\in [0,1]$ and a function $\sigma_{\rm q}(\ell) > 0$ 
defined on $]0,\ell_{\rm max}]$ such that for any sequence $L_N$ with 
$L_N / N \to \ell$, 
\begin{eqnarray}
{\rm if} \ 
\ell > \ell_{\rm max} \ , && {\rm Prob}[{\cal N}_{L_N} > 0 ] \to 0 \ ,\\
{\rm if} \  
\ell \in ]0,\ell_{\rm max}] \ , && \forall \epsilon > 0 \ \ \
{\rm Prob}\left[\left| \frac{1}{N}\log {\cal N}_{L_N} - \sigma_{\rm q}(\ell)
\right| \ge \epsilon \right] \to 0  \ .
\label{eq_cvp}
\end{eqnarray}
In the second line $\log(0)$ should be interpreted as $-\infty$, i.e. outside
of any finite interval.

The standard probabilistic methods for proving this kind of results rely 
essentially on the combinatorial computation of the average and variance of 
${\cal N}_L$, which are then used in the Markov and Chebyshev inequalities
(first and second method). The rigorous results on the number of circuits
in regular random graphs~\cite{Wormald_review,Robinson_ham,Janson_ham,Garmo} 
have indeed be obtained through a refined version of the second moment method
(see theorem 4.1 in~\cite{Wormald_review}).
In this context this approach is limited to cases where the second moment
of ${\cal N}_L$ is not exponentially larger than the square of its first 
moment\footnote{In a different problem, namely the random ensemble of 
$k$-satisfiability formulae, this limitation has been overcome by a weighted
second moment method~\cite{Achli}.}. 
The quenched entropy is then shown to be equal to the annealed one,
$\sigma_{\rm a}(\ell) = \lim \log (\overline{{\cal N}_{\ell N}}) /N$, where the
overline denotes the average over the random graph ensemble.

We believe that in all ensembles of graphs which are not strictly regular
and have a fastly decaying connectivity distribution,
the second moment of the number of long circuits is exponentially larger
that the square of the first moment (details of the combinatorial computations
leading to this belief are presented in App.~\ref{sec_app_combinatorial}), 
thus ruling out the main probabilistic techniques used so far. The annealed
entropy~\cite{BiMa} is in this case strictly larger than the quenched one, 
as it is dominated by exponentially rare graphs which have exponentially more 
circuits than the typical ones.

We shall now follow the cavity method~\cite{MePa_Bethe}, which is particularly
well suited to tackle this problem, ubiquitous in the field of disordered 
statistical mechanics models~\cite{Beyond}. 
According to this view, the quenched entropy controlling the 
leading behaviour of the number of circuits in the typical graphs depends
on the graph ensemble only through its limiting degree distribution 
$q_k$~\footnote{This is not true for the annealed entropy which depends on
the ``microscopic details'' of the ensemble. For instance the two classical
ensembles $G(N,p=c/N)$ and $G(N,M=Nc/2)$ have the same Poisson degree 
distribution but distinct annealed entropies, see Sec.~\ref{sec_enumerations}
and App.~\ref{sec_app_combinatorial}.}.
We can for instance assume that the graphs are drawn uniformly among 
all graphs on $N$ vertices which have this degree distribution.
Let us recall the existence in this case of a percolation 
transition~\cite{MR,NeStWa}
between a low connectivity regime where the
connected components of the graph are essentially trees of finite size, to
a percolated phase where one giant component contains a finite fraction
of the vertices. 

Before proceeding with the computations, we
introduce some notations used in the following.
$c=\sum_k k q_k$ 
denotes the mean connectivity of the graph, hence the
number of edges is in the thermodynamic limit $M = N c/2$.
$\tilde{q}_k$ will be the offspring probability, that is to say
the probability of finding a site of degree $k+1$ when selecting at random 
an edge of the graph and then one of its two vertices.
As a site is encountered in such a selection with a probability 
proportional to its degree, $\tilde{q}_k$ is proportional to $(k+1)q_{k+1}$. 
By normalization,
\begin{equation}
\tilde{q}_k = \frac{(k+1) q_{k+1}}{c} \ .
\label{eq_qtilde_q}
\end{equation}
To simplify notations we shall also define the factorial moments of $q_k$ and 
$\tilde{q}_k$ as
\begin{equation}
\mu_n = \sum_{k=n}^\infty q_k \, k (k-1) \dots (k-n+1) \ , \quad
\tilde{\mu}_n = \sum_{k=n}^\infty \tilde{q}_k \, k (k-1) \dots (k-n+1) \ , 
\quad
\mu_n = c \tilde{\mu}_{n-1} \ ,
\end{equation}
where the last relation is a simple consequence of Eq.~(\ref{eq_qtilde_q}).

The condition for percolation~\cite{MR,NeStWa} reads with these 
notations $\tilde{\mu}_1>1$.
We shall assume in the following that this condition is met: the long circuits
we are studying cannot be present if the graph has no giant component.

We restrict ourselves to fastly (i.e. faster than any power law) decaying 
distributions of connectivities, such that all their moments are finite.
After stating the results for 
arbitrary $q_k$ we shall often specialize to Poissonian graphs of mean 
connectivity $c$, i.e. such that $q_k = e^{-c} c^k/ k!$.

\subsection{The quenched computation}
\label{sec_quenched}
In essence the computation of the quenched entropy we undertake now
amounts to perform the Bethe approximation of the statistical model
defined by Eqs.~(\ref{eq_law},\ref{eq_wi}) for graphs generated according to
the connectivity distribution $q_k$. The solution of the BP equations
(\ref{eq_msgs}), which depends on the particular graph on which they are
applied, leads then to a random set of messages $y$. Taking at random
a graph of the ensemble, and a directed edge of this graph, one finds a 
message $y$ with probability distribution $P(y;u)$. 
In the so-called cavity 
method at the replica-symmetric level~\cite{MePa_Bethe}, one assumes that
the incoming messages on this directed edge are independent random variables
with the same probability law $P(y;u)$. Using Eq.~(\ref{eq_msgs}), this is
turned into a self-consistent equation,
\begin{equation}
P(y;u) = \tilde{q}_0 \, \delta(y) + \sum_{k=1}^\infty \tilde{q}_k  
\int_0^\infty dy_1 \, P(y_1;u) \dots
dy_k \, P(y_k;u) \, \delta(y-g_k(y_1,\dots,y_k)) \ ,
\label{eq_Py}
\end{equation}
where we have defined
\begin{equation}
g_1(y_1) = u y_1 \ , \qquad g_k(y_1,\dots,y_k)= 
\frac{u \overset{k}{\underset{i=1}{\sum}} y_i }
{1+ u^2 \underset{1\le i <j \le k}{\sum} y_i y_j } \ 
{\rm for} \ k\ge 2 \ . 
\label{eq_def_f}
\end{equation}

The quenched free-energy is then expressed in terms
of this $P(y;u)$ as (cf. Eq.~(\ref{eq_f})):
\begin{eqnarray}
f_{\rm q}(u) &=& \sum_{k=2}^\infty q_k \int_0^\infty dy_1 \, P(y_1;u) \dots
dy_k \, P(y_k;u) \, \ln \left( 1 + u^2 \sum_{1\le i < j \le k} y_i y_j\right)
\nonumber \\ &\phantom{=}& 
-\frac{c}{2} \int_0^\infty dy_1 \, P(y_1;u) \, dy_2 \, P(y_2;u) \, 
\ln (1+ u \,y_1 y_2)
\ , \label{eq_fbeta}
\end{eqnarray}
In a similar way the length of the circuits in the configurations selected by 
a given value of $u$, and the corresponding quenched entropy read:
\begin{eqnarray}
\ell(u)&=& \frac{c}{2} \int_0^\infty dy_1 \, P(y_1;u) \, dy_2 \, P(y_2;u) \,
\frac{u y_1 y_2}{1+u y_1 y_2} \ , \label{eq_lbeta} \\
\sigma_{\rm q}(\ell(u)) &=& f_{\rm q}(u) - \ell(u) \ln u \ .
\label{eq_sigmabeta}
\end{eqnarray}

As appears clearly when considering Eq.~(\ref{eq_Py}), the distribution 
$P(y;u)$ contains a Dirac's delta in $y=0$, that is to say a finite fraction 
of the messages are strictly vanishing.
Let us call $\eta$ the fraction of non-trivial 
messages\footnote{$\eta$ was denoted $1-\zeta$ in~\cite{EPL}.}, and 
$\widehat{P}(y;u)$ their (normalized) distribution, i.e.
$P(y;u) = (1-\eta) \, \delta(y) + \eta \, \widehat{P}(y;u)$
where $\widehat{P}$ does not contain a Dirac's delta in $y=0$. Inserting this 
definition in Eq.~(\ref{eq_Py}), one obtains the equation satisfied by
$\eta$,
\begin{equation}
1-\eta = \sum_{k=0}^\infty \tilde{q}_k (1-\eta)^k \ .
\label{eq_eta}
\end{equation}
Besides the trivial solution $\eta=0$, this equation has another positive
solution as soon as $\tilde{\mu}_1 > 1$, i.e. when the graph is in the
percolating regime. One also realizes that $\widehat{P}$ satisfies the 
equation 
obtained from Eq.~(\ref{eq_Py}) by replacing the offspring distribution 
$\tilde{q}$ by $\tilde{r}$, defined as
\begin{equation}
\tilde{r}_0 = 0 \ , \qquad
\tilde{r}_k =\sum_{n=k}^\infty \tilde{q}_n {n \choose k} \eta^{k-1} 
(1-\eta)^{n-k} \quad {\rm for} \ k \ge 1 \ .
\label{eq_rtilde}
\end{equation}
Finally the free-energy and the typical length of circuits 
(cf. Eqs.~(\ref{eq_fbeta},\ref{eq_lbeta}))
can also be expressed in terms of the simplified distribution $\widehat{P}$,
if one replaces $q$ by the following distribution $r$:
\begin{equation}
r_0 = 1 - \sum_{k \ge 2} r_k \ , \quad r_1=0 \ , \quad
r_k = \sum_{n=k}^\infty q_n {n \choose k} \eta^k (1-\eta)^{n-k} 
\quad {\rm for} \ k \ge 2 \ .
\label{eq_r}
\end{equation}
It is easily verified that this modified distribution has mean $c \eta^2$,
and that a relation similar to Eq.~(\ref{eq_qtilde_q}) holds between $r$ and
$\tilde{r}$,
\begin{equation}
\tilde{r}_k = \frac{(k+1)r_{k+1} }{c \eta^2} \ .
\end{equation}
Let us now give the interpretation of this simplification process. We have
shown the equality of the quenched entropy of the circuits in the two ensembles
defined one by $q_k$, the other by $r_k$. As we explained 
at the end of Sec.~\ref{sec_bethe}, the circuits of a graph $G$ necessarily 
belong to its 2-core, that is to say the largest subgraph of $G$ in which all
vertices have a degree at least equal to two. On the dangling ends, i.e. 
the edges that do not belong to the 2-core, at least one of the two
directed messages $y$ is equal to zero. It is thus very natural to
interpret the elimination of null messages in terms of the typical properties
of the 2-core of graphs drawn from the ensemble defined by the 
distribution $q_k$.
The fraction of edges in the 2-core should be $\eta^2$, as both directed 
messages have to be non-zero for the edge to belong to the 2-core, $r_k$
(resp. $\tilde{r}_k$) should be the connectivity (resp. offspring) distribution
of the 2-core. This interpretation is indeed confirmed by
a direct study of a leaf-removal algorithm which iteratively removes the
dangling ends of a graph, that we present in App.~\ref{sec_app_leaf}. In the
following we shall use the distribution $q$ or $r$, depending on which is 
simpler in the encountered context.

For future use we give the explicit expressions in the case of Poissonian 
random graphs,
\begin{equation}
\eta = 1- e^{-c \eta} \ , \qquad 
r_k = \frac{e^{-c \eta}(c\eta)^k}{k!} \ \ \ {\rm for} \ \  k \ge 2 \ , \qquad 
\tilde{r}_k = \frac{1}{\eta} \frac{e^{-c \eta}(c\eta)^k}{k!} \ \ \
{\rm for} \ \ k \ge 1 
\ .
\label{eq_r_poisson}
\end{equation}

We now come back to the predictions of the quenched entropy and
consider as a first example the case of regular graphs of
connectivity $c$, for which $\tilde{q}_k=\delta_{k,c-1}$.
Equation (\ref{eq_Py}) on the distribution of messages has then a very simple
solution, $P(y;u)=\delta(y-y_{\rm r}(u,c))$, with
\begin{equation}
y_{\rm r}(u,c)=\sqrt{\frac{2 u(c-1)-2}{u^2 (c-1)(c-2)}} \ .
\label{eq_y_regular}
\end{equation}
It is then straightforward to express $\ell(u)$ and 
$\sigma_{\rm q}(\ell(u))$ from this solution. One can also eliminate the 
parametrization by $u$ to obtain the entropy,
\begin{equation}
\sigma_{\rm r}(\ell,c) 
= -(1-\ell) \ln(1-\ell) + \left( \frac{c}{2} - \ell \right) 
\ln \left( 1 - \frac{2 \ell}{c} \right) + \ell \ln (c-1) \ ,
\label{eq_sigma_regular}
\end{equation}
which corresponds to the known results mentioned 
above~\cite{Garmo,MaMo}\footnote{These results are also a particular
case of the study of polymers on regular graphs presented in~\cite{polymers}.}.

The peculiarity of the regular case for which annealed and quenched averages
coincide is hinted to by the simplicity of this solution $P(y;u)$ with a
single Dirac peak. As soon as $\tilde{q}_k$ is positive for more than one
connectivity, the distribution $P(y;u)$ 
acquires a non vanishing support, which
we expect to show up as larger fluctuations in the numbers of circuits, 
and hence a difference between quenched and annealed computations.

\begin{figure}
\begin{center}
\includegraphics[width=9cm]{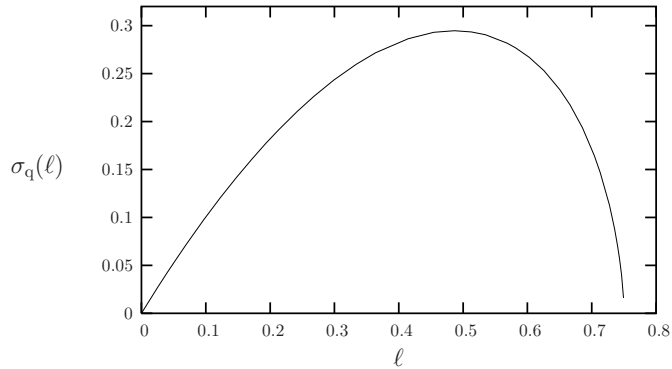}
\end{center}
\caption{Quenched entropy for a Poissonian graph of mean connectivity 3.}
\label{fig_ex_poisson}
\end{figure}

The equation on $P(y;u)$ is not solvable analytically for an arbitrary 
connectivity distribution.
Two complementary roads can then be followed:
this distributional equation can be easily solved with a population dynamics
algorithm~\cite{MePa_Bethe}. One represents $P(y;u)$ by a sample of a large
number of $y$'s, at each time step one draws a number $k$ following the
law $\tilde{r}_k$, extracts $k$ values $y_1,\dots,y_k$ randomly from the 
population, computes the new value $g_k(y_1,\dots,y_k)$, and replace one
of the representant of the population by this new value.
Starting from a random sample of messages, the population 
converges to a sample of message distributed according to the fixed point 
solution of Eq.~(\ref{eq_Py}). The corresponding physical observables are
then computed from this sample of messages, which yields the prediction 
for $\sigma_{\rm q}(\ell)$. As an illustration, we present in 
Fig.~\ref{fig_ex_poisson} the results of such a numerical computation for 
Poissonian graphs of mean degree $c=3$.

On the analytical side, we present in
the next sections a study of two limits, for the short
and longest circuits, in which the analytical predictions can be pushed
forward. 

\section{The limit of small circuits}

We shall investigate in this section the behaviour of the quenched
entropy in the limit of small circuits, computing analytically its first
two derivatives at the origin, $\sigma_{\rm q}'(0)$ and $\sigma_{\rm q}''(0)$.

Let us first show the existence of a critical value $u_{\rm m}$ below
which the typical configurations are deprived of any circuit. This transition
is signaled by an instability of the trivial solution of Eq.~(\ref{eq_Py}), 
$P(y)=\delta(y)$. Perturbing this distribution infinitesimally, 
one can expand $g_k$ as 
$g_k(y_1,\dots,y_k)= u \sum_{i=1}^k y_i + O(y_i^3)$. In this limit, if
one inserts in the r.h.s. of Eq.~(\ref{eq_Py}) some $P(y)$ with an 
infinitesimal mean $P_1$, one obtains another distribution with a mean
$(u/u_{\rm m}) P_1$, with
\begin{equation}
u_{\rm m}^{-1} = \sum_{k=1}^\infty k \tilde{q}_k = \tilde{\mu}_1 \ .
\end{equation}
If $u< u_{\rm m}$, the mean of the perturbed distributions decrease 
upon iteration, so $P(y) = \delta(y)$ is a stable solution. 
On the contrary if $u> u_{\rm m}$ this solution is unstable and must 
flow to a non-trivial stable fixed point. 

We shall now set up an expansion around the stability limit, 
$u = u_{\rm m} + \epsilon$ with $\epsilon \to 0^+$. In this
limit the messages $y$ are supported on a scale which vanishes with $\epsilon$,
we shall consequently define $y=x \, \epsilon^a + o(\epsilon^a)$ with $x$ 
finite, and $a$ a positive exponent to be determined in a few lines. Let us 
denote $Q(x)$ the distribution of the rescaled messages and
relate the moments of $P$ and $Q$ as
\begin{equation}
P_n(\epsilon) = \int_0^\infty dy \, y^n P(y;u_{\rm m}+ \epsilon) \sim 
\epsilon^{na} \int_0^\infty dx \, x^n Q(x) = \epsilon^{na} Q_n \ .
\end{equation}
Expanding Eq.~(\ref{eq_def_f}) as
\begin{equation}
x \sim u_{\rm m} \sum_{i=1}^k x_i + \epsilon \sum_{i=1}^k x_i
- u_{\rm m}^2 \epsilon^{2 a} \left( \sum_{i=1}^k x_i \right)
\left( \sum_{1\le i<j \le k} x_i x_j \right) \ ,
\end{equation}
we obtain at the lowest order
\begin{eqnarray}
\epsilon &=& u_{\rm m}^4 \tilde{\mu}_2 Q_2 \epsilon^{2a} 
+ \frac{1}{2} u_{\rm m}^4 \tilde{\mu}_3 Q_1^2 \epsilon^{2a} \ , \\
Q_2 \epsilon^{2a} &=& u_{\rm m} Q_2 \epsilon^{2a} 
+ u_{\rm m}^2 \tilde{\mu}_2 Q_1^2 \epsilon^{2a} \ .
\end{eqnarray}
This fixes the scale $a=1/2$ and the values of $Q_1$ and $Q_2$, the
first two moments of the distribution solution of
\begin{equation}
Q(x) = \tilde{q}_0 \, \delta(x) + \sum_{k=1}^\infty \tilde{q}_k 
\int_0^\infty dx_1 Q(x_1) \dots dx_k Q(x_k) \, 
\delta\left(x - u_{\rm m} \sum_{i=1}^k x_i\right) 
\ .
\label{eq_Qx}
\end{equation}

Let us now consider the consequences of this scaling on the observables 
$f_{\rm q}$, $\ell$, $\sigma_{\rm q}$ in the limit $u \to u_{\rm m}$. 
Taking into account both their explicit dependence on $u$ and the implicit one
through the distribution $P(y;u)$, one finds after a short computation that:
\begin{itemize}
\item[$\bullet$]
the expansion of $f_{\rm q}(u_{\rm m} + \epsilon)$ starts at the second order,
\begin{eqnarray}
f_{\rm q}(u_{\rm m} + \epsilon) &=& 
f_{\rm q}^{(2)} \epsilon^2 + O(\epsilon^3) \ , \\
f_{\rm q}^{(2)} &=& 
\frac{c}{2} Q_1^2 + 
\frac{c}{4} u_{\rm m}^2 (1 - u_{\rm m}) Q_2^2
- \frac{1}{2} u_{\rm m}^4 \mu_3 Q_1^2 Q_2
- \frac{1}{8} u_{\rm m}^4 \mu_4 Q_1^4 \ .
\end{eqnarray}
\item[$\bullet$] 
the intensive length $\ell$ is, at its first non-trivial order,
\begin{equation}
\ell(u_{\rm m} + \epsilon) = \ell^{(1)} \epsilon + O(\epsilon^2) \ ,
\qquad \ell^{(1)} = \frac{c}{2} u_{\rm m} Q_1^2 \ .
\end{equation}
\item[$\bullet$]
the first two derivatives of $\sigma_{\rm q}(\ell)$ in $\ell=0$ can be 
obtained from the previous expressions:
\begin{equation}
\sigma_{\rm q}'(0) = - \ln u_{\rm m} \ , \qquad
\sigma_{\rm q}''(0) = \frac{2 f_{\rm q}^{(2)}}{(\ell^{(1)})^2} - 
\frac{2}{u_{\rm m} \ell^{(1)}}
\label{eq_derivsq}
\end{equation}
\end{itemize}
Solving for $Q_{1,2}$ and plugging their values in the expression 
(\ref{eq_derivsq}) of the derivatives 
of the entropy one finally obtains:
\begin{equation}
\sigma_{\rm q}'(0) = \ln \tilde{\mu}_1 \ , \qquad
\sigma_{\rm q}''(0) = - \frac{1}{c} \left( 
\frac{\tilde{\mu}_3 }{\tilde{\mu}_1^2}
+ \frac{2 \tilde{\mu}_2^2 }{\tilde{\mu}_1^3 (\tilde{\mu}_1 -1)} \ .
\right)
\label{eq_short}
\end{equation}

We can now turn to the discussion of these results, and in particular
to the comparison with the annealed computation of Bianconi and 
Marsili~\cite{BiMa}. Expanding their result 
(reproduced in Eq.~(\ref{eq_sigmaa})) in
powers of $\ell$, one obtains
\begin{equation}
\sigma_{\rm a}'(0) = \ln \tilde{\mu}_1 \ , \qquad
\sigma_{\rm a}''(0) = -\frac{1}{c} 
\frac{\tilde{\mu}_3 + 4 \tilde{\mu}_2 - 2 \tilde{\mu}_1 (\tilde{\mu}_1 -1)}
{\tilde{\mu}_1^2} \ .
\end{equation}
Let us first consider a large but non extensive circuit length, 
$1\ll L \ll \ln N$.
The number ${\cal N}_L$ of such circuits is, in the thermodynamic limit, 
a Poisson distributed random variable with a mean equal to
\begin{equation}
\frac{1}{2L}\left( \frac{\sum_k k(k-1) q_k }{\sum_k k q_k }\right)^L = 
\frac{1}{2L}(\tilde{\mu}_1)^L \ .
\end{equation}
When $L \gg 1$ the most probable value of this random variable is equal to
its mean, in which one can neglect the polynomial prefactor $1/(2L)$ (we
recall that we assume $\tilde{\mu}_1>1$ to be in the percolated regime).
Consequently the quenched and annealed computation of the first derivative
of the entropy at $\ell=0$ coincide and match the result for 
$1\ll L \ll \ln N$:
\begin{equation}
{\cal N}_L \sim e^{N \sigma(L/N)} \sim e^{L \sigma'(0)} = 
(\tilde{\mu}_1 )^L \ .
\end{equation}
On the contrary the second derivatives differ, in general, in the two 
computations:
\begin{equation}
\sigma_{\rm a}''(0) - \sigma_{\rm q}''(0) = 
\frac{2}{c \tilde{\mu}_1^3 (\tilde{\mu}_1-1)} (\tilde{\mu}_2 - 
\tilde{\mu}_1 (\tilde{\mu}_1 -1))^2 \ .
\end{equation}
As expected 
the annealed entropy is always greater than the quenched one at this order of 
the expansion. Moreover it is straightforward to show from the above
expression that $\sigma_{\rm a}''(0) - \sigma_{\rm q}''(0)$ vanishes only if
the distribution $\tilde{q}_k$ is supported by a single integer, in other
words in the random regular graph case. 

We performed this computation using the degree distribution $q_k$ of the graph,
however the reader will easily verify that Eq.~(\ref{eq_short}) remains
unchanged if one replaces $q_k$ by the connectivity distribution $r_k$ of
its 2-core (the factorial moments $\tilde{\mu}_n$ gets multiplied by
$\eta^{n-1}$, the mean connectivity $c$ by $\eta^2$).

For completeness we state here the results for Poissonian graphs of
average degree $c$,
\begin{equation}
\sigma_{\rm q}'(0) = \ln c \ , \quad \sigma_{\rm q}''(0) = - \frac{c+1}{c-1} \
.
\end{equation}

\section{The limit of longest circuits}
A more interesting limit case is the one of maximal length circuits.
Some questions
arise naturally in this context: what is the maximal length, $\ell_{\rm max}$,
for which circuits of $N \ell_{\rm max}$ edges are present with high
probability in a given random graph ensemble?
In particular, under which conditions these graphs are Hamiltonian, that is
to say $\ell_{\rm max}=1$?
Finally, what is the number of such longest 
circuits, measured by the corresponding quenched entropy 
$\sigma_{\rm q}(\ell_{\rm max})$? From the properties of the Legendre 
transform (cf. Eq.~(\ref{eq_Legendre})), these quantities can be determined
by investigating the limit $u \to \infty$ of the free-energy~:
\begin{equation}
f_{\rm q}(u) \underset{u \to \infty}{\sim} \ell_{\rm max} \ln u
+ \sigma_{\rm q}(\ell_{\rm max}) \ .
\label{eq_f_betatoinf}
\end{equation}
This corresponds, in the jargon of the statistical mechanics approach to 
combinatorial optimization problems, to a zero temperature limit, where
$-\ell_{\rm max}$ (resp. $\sigma_{\rm q}(\ell_{\rm max})$) is the
ground-state energy (resp. entropy) density. 

It turns out that the answers to the above questions crucially depend on the
 presence or not of degree 2 sites in the 2-core of the random graphs under 
study, we shall thus divide the rest of this section according to 
this distinction. Before that we state an equivalent expression of the 
free-energy which will prove more convenient in this limit,
\begin{equation}
f_{\rm q}(u)=
\sum_{k=3}^\infty q_k \left( \frac{k}{2} I_{k-1}(u) 
- \frac{k-2}{2} I_k(u) \right) \ ,
\label{eq_fbeta_I}
\end{equation}
where we have defined some logarithmic moments of the distribution $P$,
\begin{equation}
I_k(u)=\int_0^\infty dy_1 \, P(y_1;u) \dots dy_k \, P(y_k;u) \ln
\left(1 + u^2  \underset{1\le i<j \le k}{\sum} y_i y_j \right)
\quad {\rm for} \ k \ge 2. 
\label{eq_Ibeta}
\end{equation}
This form of $f_{\rm q}(u)$ is obtained from Eq.~(\ref{eq_fbeta}) by using
the equation (\ref{eq_Py}) on $P(y;u)$ and the identity
\begin{equation}
1+ u \, y_0 \, g_k(y_1,\dots,y_k) = 
\frac{1 + u^2  \underset{0\le i<j \le k}{\sum} y_i y_j }
{1 + u^2  \underset{1\le i<j \le k}{\sum} y_i y_j} \ .
\end{equation}

\subsection{In the absence of degree 2 sites in the 2-core}
\label{sec_binf_without}
One can gain some intuition on the limit $u \to \infty$ by inspecting
the behaviour of the messages in the regular case. Indeed, the expansion
of Eq.~(\ref{eq_y_regular}) shows a scaling of the form 
$y \sim x u^{-1/2}$, with $x$ (an evanescent field in the jargon
of optimization problems) finite. Consider now the more general case
of random graph ensembles with a minimal connectivity of 3, i.e.
$q_0=q_1=q_2=0$, which consequently implies $\tilde{q}_0=\tilde{q}_1=0$. Thanks
to the vanishing of $\tilde{q}_1$, the equations (\ref{eq_Py},\ref{eq_def_f})
have a solution with the above scaling of $y$ with $u$. One can also 
check numerically in particular cases that the distributions $P(y)$ concentrate
according to this behaviour for large but finite values of $u$.

Denoting $V_0(x)$ the distribution of the evanescent fields, one easily
obtains the integral equation it obeys:
\begin{equation}
V_0(x) = \sum_{k=2}^\infty \tilde{q}_k  \int_0^\infty dx_1 \, V_0(x_1) \dots
dx_k \, V_0(x_k) \, \delta(x-h_k(x_1,\dots,x_k)) \ , \quad {\rm with} \ \ 
h_k(x_1,\dots,x_k)= 
\frac{\overset{k}{\underset{i=1}{\sum}} x_i }
{\underset{1\le i <j \le k}{\sum} x_i x_j } \ . 
\label{eq_Rx}
\end{equation}
Moreover the logarithmic moments defined
in Eq.~(\ref{eq_Ibeta}) have the following scaling in this limit,
\begin{equation}
I_k(u) \sim \ln u + J_k \ , \quad {\rm with} \ 
J_k=\int_0^\infty dx_1 \, V_0(x_1) \dots dx_k \, V_0(x_k) \ln
\left( \underset{1\le i<j \le k}{\sum} x_i x_j \right) \ .
\end{equation}
Plugging this equivalent in the expression (\ref{eq_fbeta_I}) for the
quenched free-energy, and identifying
the maximal length of the circuits with the coefficient of order $\ln u$,
and their entropy with the constant term, we obtain:
\begin{equation}
\ell_{\rm max}=1 \ , \qquad \sigma_{\rm q}(1) = 
\sum_{k=3}^\infty q_k \left( \frac{k}{2} J_{k-1} - \frac{k-2}{2} J_k \right) 
\ .
\label{eq_Wormald_refined}
\end{equation}

\begin{figure}
\begin{center}
\includegraphics[width=9cm]{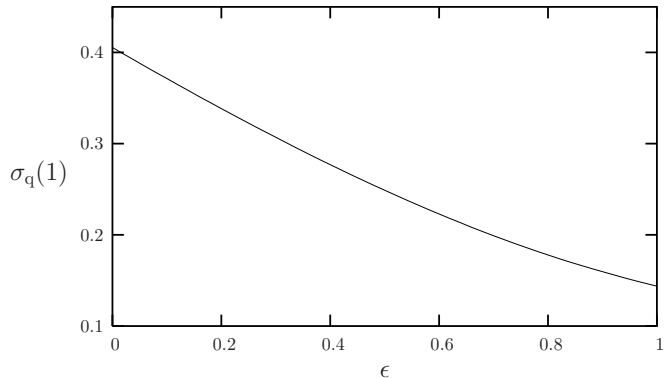}
\end{center}
\caption{Quenched entropy for the Hamiltonian circuits in graphs with a
fraction $\epsilon$ of degree 3 vertices, and $1-\epsilon$ of degree 4.}
\label{fig_ex_3p4_ent_hamiltonian}
\end{figure}

The identity $\ell_{\rm max}=1$ (of which we present an alternative derivation
in App.~\ref{app_lmax}) reproduces the conjecture of Wormald 
(conjecture 2.27 in~\cite{Wormald_review}) that 
random graphs with a minimum connectivity of 3 are, with high probability,
Hamiltonian.
Obviously the methods we used are far from rigorous and do not provide a
valid proof of the conjecture. However they give it a quantitative flavour
with the prediction of $\sigma_{\rm q}(1)$, the typical entropy of such 
Hamiltonian circuits.
We performed a numeric resolution of Eq.~(\ref{eq_Rx}), again by a
population dynamics algorithm, to compute the moments $J_k$ and from them
the quenched entropy $\sigma_{\rm q}(1)$. As an illustrative example,
Fig.~\ref{fig_ex_3p4_ent_hamiltonian} presents the results of such a 
computation in the case of random graphs
with a fraction $\epsilon$ of degree 3 vertices, and $1-\epsilon$ of degree
4. As a function of $\epsilon$ the entropy interpolates between the 
rigorously known values at $\epsilon=0$, $\epsilon=1$, for which the graphs
are regular. Note that the quenched and annealed entropies, even if 
strictly different when $0<\epsilon<1$, are found to be numerically close.
For instance when $\epsilon=0.5$, one has $\sigma_{\rm q}(1) \approx 0.2489$ 
and $\sigma_{\rm a}(1) \approx 0.2501$.

\subsection{In the presence of degree 2 sites in the 2-core}
\label{sec_binf_with}

Let us now consider the question of the longest cycles in random graph 
ensembles with connectivity distribution $q_k$ which do not fulfill the
condition $q_0=q_1=q_2=0$ we assumed in the previous subsection. We first
present simple combinatorial arguments which lead to bounds on $\ell_{\rm max}$
and to an asymptotic expansion when there are very few degree 2 sites in
the 2 core, before coming back to the limit $u \to \infty$ of the cavity 
approach.

\subsubsection{Bounds on $\ell_{\rm max}$}
\label{sec_bounds}
Let us call $G_1$ a graph drawn at random from such an ensemble, and 
$G_2$ its 2-core, determined for instance by the leaf removal algorithm 
detailed in App.~\ref{sec_app_leaf}. $G_2$ has
the connectivity distribution $r_k$ defined and computed in 
Sec.~\ref{sec_quenched} and App.~\ref{sec_app_leaf}. The number of sites in the
2-core is $N \ell_{\rm core}$ (cf. Eq.~\ref{eq_lcore}), with 
$\ell_{\rm core}<1$ unless the original graph was deprived of any isolated
sites and of leaves (i.e. $q_0=q_1=0$). This $\ell_{\rm core}$ is clearly
an upper bound on $\ell_{\rm max}$, as circuits cannot be longer that the
number of available sites in the 2-core.

\begin{figure}
\includegraphics[width=7cm]{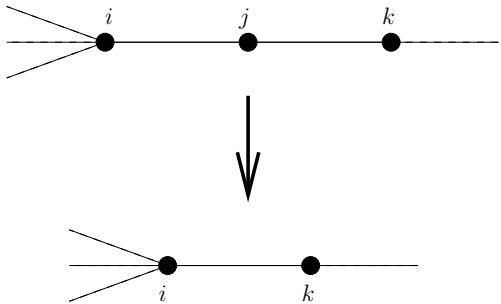}
\caption{A single step of the simplification algorithm (from $G_2$ to $G_3$)
used in Sec.~\ref{sec_binf_with}.}
\label{fig_elim_chain}
\end{figure}

One can derive a lower bound on $\ell_{\rm max}$ with the following
reasoning. From $G_2$, the 2-core of $G_1$, eliminate
recursively sites of degree 2, identifying the two edges which were 
previously incident to it (see Fig.~\ref{fig_elim_chain}). When all sites
of degree 2 have been removed, one ends up with a graph, call it $G_3$, on 
$N (\ell_{\rm core} -r_2)$ sites, where the minimal connectivity is 3. Using
the result of the previous section, this reduced graph typically contains 
circuits of length $N (\ell_{\rm core} -r_2)$. Each of the circuits of
$G_3$ can be unambiguously associated to a circuit of $G_2$, reinserting
the edges which were simplified during the construction of $G_3$. Obviously
the reconstructed circuits of $G_2$ are longer than the ones of $G_3$,
hence $l_{\rm lb}=(\ell_{\rm core} -r_2)$ should be a lower bound
for $\ell_{\rm max}$. These bounds have been used under a stronger form in 
the case
of Erdos-Renyi random graphs very close to the percolation threshold 
(for $c=1+\delta$ with $N^{-1/3}\ll \delta \ll 1$) in~\cite{Luczak}.

One can then wonder if the upper bound is
saturated, in other words if the 2-core is Hamiltonian. In general the answer
is no, as explained by the following remark. Consider a site of degree 
strictly greater than 2, surrounded by at least three neighbors of degree 2: 
obviously, no circuit can go through more than two of these neighbors.
As soon as $r_2 >0$ there will be an extensive number of such
forbidden vertices, hence in such a case 
$\ell_{\rm max}<\ell_{\rm core}$. The equality is possible only if $r_2=0$,
which was the case investigated in the previous section.

Note that the gap between the lower and upper bound closes when $r_2$ vanishes,
as the 2-core becomes Hamiltonian in this limit. A conjecture on the
behaviour of $\ell_{\rm max}$ in the limit $r_2 \to 0$ can be formulated as
\begin{equation}
\ell_{\rm max} = \ell_{\rm core} - \sum_{k=3}^\infty r_k {k \choose 3} 
\tilde{r}_1^3 + O(\tilde{r}_1^4) \ .
\label{eq_lmax_expansion}
\end{equation}
This expression has a very simple interpretation: a forbidden site in the above
argument appears if a vertex of degree greater than 3 is surrounded by at
least three vertices of degree 2. At the lowest order these forbidden sites
are far apart from each other, $N \ell_{\rm max}$ is thus reduced by one
unit each time this appears. This conjecture will come out of the cavity
analysis of next subsection, we preferred to anticipate it here because of
its simple combinatorial interpretation.

Let us exemplify the bounds and the conjecture in two particular cases.
Consider ensemble of graphs where a fraction $1-\epsilon$ of vertices
have degree $c_0 \ge 3$, the others being of degree 2, 
with $0\le \epsilon<1$. The above bounds and estimation read
\begin{equation}
1 - \epsilon \le \ell_{\rm max} \le 1 \ , \quad 
\ell_{\rm max} = 1 - \frac{4 (c_0-1)(c_0-2)}{3 c_0^2} \epsilon^3 
+ O(\epsilon^4) \ .
\end{equation}

As a second example we consider Poissonian random graphs of mean degree $c$,
for which the bounds read (cf. Eq.~(\ref{eq_r_poisson}))
\begin{equation}
\ell_{\rm lb} = 1 - (1-\eta)\left(1+c\eta+ \frac{1}{2}(c\eta)^2\right) 
\ , \qquad \ell_{\rm core} = 1 - (1-\eta)(1+c\eta) \ ,
\label{eq_bounds_poissonian}
\end{equation}
where $\eta$ is the solution of $\eta = 1- \exp(-c \eta)$.
In the limit $c \to \infty$ the fraction of degree 2 vertices in the 2-core
vanishes, the above conjecture reads then
\begin{equation}
\ell_{\rm max} = 1 - (c+1) e^{-c} - c^2 e^{-2c} - \frac{c^2}{2}
\left( \frac{c^4}{3} + 3 c -1 \right) e^{-3c} + O(c^n e^{-4c} ) \ ,
\end{equation}
where $n$ is some positive integer. Most of these terms come from
the expansion of $\ell_{\rm core}$, the only non-trivial one being 
$-c^6 e^{-3c}/6$. This is in agreement with a rigorous result of 
Frieze~\cite{Frieze},
\begin{equation}
1-(1 + \delta(c)) c e^{-c}  \le  \ell_{\rm max} \le 1 - (c+1) e^{-c} \ \quad
{\rm with} \ \ \delta(c) \underset{c \to \infty}{\to} 0 \ .
\end{equation}

\subsubsection{The large $u$ limit in presence of sites of degree 2}

We come back to the cavity approach and investigate the large
$u$ limit in this case. The simple ansatz $y \sim x u^{-1/2}$
is not compatible with Eqs.~(\ref{eq_Py},\ref{eq_def_f}) any longer,
because of the non vanishing value of $\tilde{r}_1$. The most natural
generalization which allows to close this set of equations is then
$y \sim x u^{p-1/2}$, where $p$ is a relative integer (hard field). 
We denote $V_p(x)$ the
probability distribution of the evanescent fields $x$ associated to
hard fields $p$. For notational simplicity we take the $V_p$ to be
unnormalized, with $\int_0^\infty dx \, V_p(x)= v_p$, and impose
the condition $\sum_{p\in \mathbb{Z}} v_p =1$.
Expanding Eqs.~(\ref{eq_Py},\ref{eq_def_f}) with this ansatz, one finds
\begin{equation}
V_p(x) = \sum_{k=1}^\infty \tilde{r}_k 
\sum_{p_1,\dots,p_k \in \mathbb{Z}^k} \delta_{p,e_k(p_1,\dots,p_k)}
\int_0^\infty dx_1 \, V_{p_1}(x_1) \dots dx_k \,  V_{p_k}(x_k) \,
\delta(x- h_k(p_1,x_1,\dots,p_k,x_k)) \ .
\label{eq_Vpx}
\end{equation}
In order to simplify the expression of $e_k$ and $h_k$, we shall denote
$[n]$ a permutation of the indices which orders the hard fields in
decreasing order, $p_{[1]} \ge p_{[2]} \ge \dots p_{[k]}$. 
Then
\begin{equation}
e_1(p_1) = 1+p_1 \ , \quad
e_k(p_1,\dots,p_k) = \min (1 + p_{[1]},-p_{[2]}) \quad {\rm for} \ k \ge 2 \ .
\end{equation}
We also define
\begin{eqnarray}
d_k(p_1,x_1,\dots,p_k,x_k) &=& \begin{cases} 
1 & {\rm if} \ p_{[1]} + p_{[2]} < -1 \\
1 + \hat{d}_k(p_1,x_1,\dots,p_k,x_k) & {\rm if} \ p_{[1]} + p_{[2]} = -1 \\
\hat{d}_k(p_1,x_1,\dots,p_k,x_k) & {\rm if} \ p_{[1]} + p_{[2]} > -1
\end{cases}
\ , \\
\hat{d}_k(p_1,x_1,\dots,p_k,x_k) &=& \begin{cases}
\underset{i<j | p_i=p_j=p_{[1]}}{\sum} x_i x_j & {\rm if} \ p_{[1]}=p_{[2]} \\
x_{[1]} \underset{i | p_i=p_{[2]}}{\sum} x_i & {\rm if} \ p_{[1]} > p_{[2]}
\end{cases} 
\ ,
\end{eqnarray}
in terms of which the evanescent contribution reads 
\begin{equation}
h_1(p_1,x_1)=x_1 \ , \quad h_k(p_1,x_1,\dots,p_k,x_k) = 
\frac{\underset{i | p_i = p_{[1]}}{\sum} x_i}{d_k(p_1,x_1,\dots,p_k,x_k)}
\quad {\rm for} \ k \ge 2 \ .
\end{equation}
Within this ansatz the logarithmic moments (cf. Eq.~(\ref{eq_Ibeta})) behave
as
\begin{equation}
I_k(u) \sim J_k^{({\rm h})} \ln u + J_k \ ,
\end{equation}
\begin{equation}
J_k^{({\rm h})} = \underset{p_1,\dots,p_k \in \mathbb{Z}^k}{\sum}
v_{p_1} \dots v_{p_k} \max(1+p_{[1]}+p_{[2]},0) \ ,
\label{eq_Jkh}
\end{equation}
\begin{equation}
J_k = \underset{p_1,\dots,p_k \in \mathbb{Z}^k}{\sum}
\int_0^\infty dx_1 \, V_{p_1}(x_1) \dots dx_k \,  V_{p_k}(x_k) \,
\ln d_k(p_1,x_1,\dots,p_k,x_k) \ .
\label{eq_Jk}
\end{equation}
From the behaviour of $f_{\rm q}$ we thus obtain
\begin{equation}
\ell_{\rm max}=\sum_{k=3}^\infty r_k 
\left( \frac{k}{2} J_{k-1}^{({\rm h})} - \frac{k-2}{2} J_k^{({\rm h})} \right) 
\ , \qquad 
\sigma_{\rm q}(\ell_{\rm max}) = \sum_{k=3}^\infty r_k 
\left( \frac{k}{2} J_{k-1} - \frac{k-2}{2} J_k \right) \ .
\label{eq_lmax}
\end{equation}
Obviously this set of expressions reduces to the ones of 
Sec.~\ref{sec_binf_without} when all the hard fields are null, which is a
solution if and only if $\tilde{r}_1=0$. 

\begin{figure}
\begin{center}
\includegraphics[width=8cm]{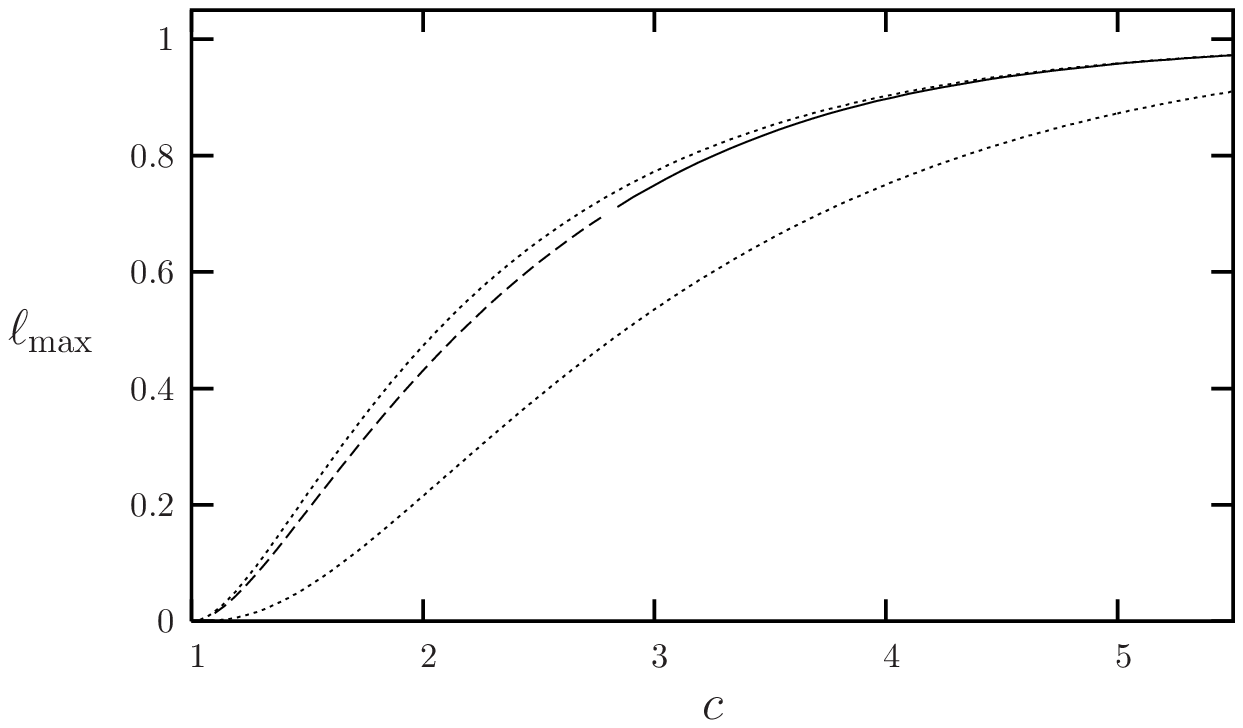} \hspace{7mm}
\includegraphics[width=8cm]{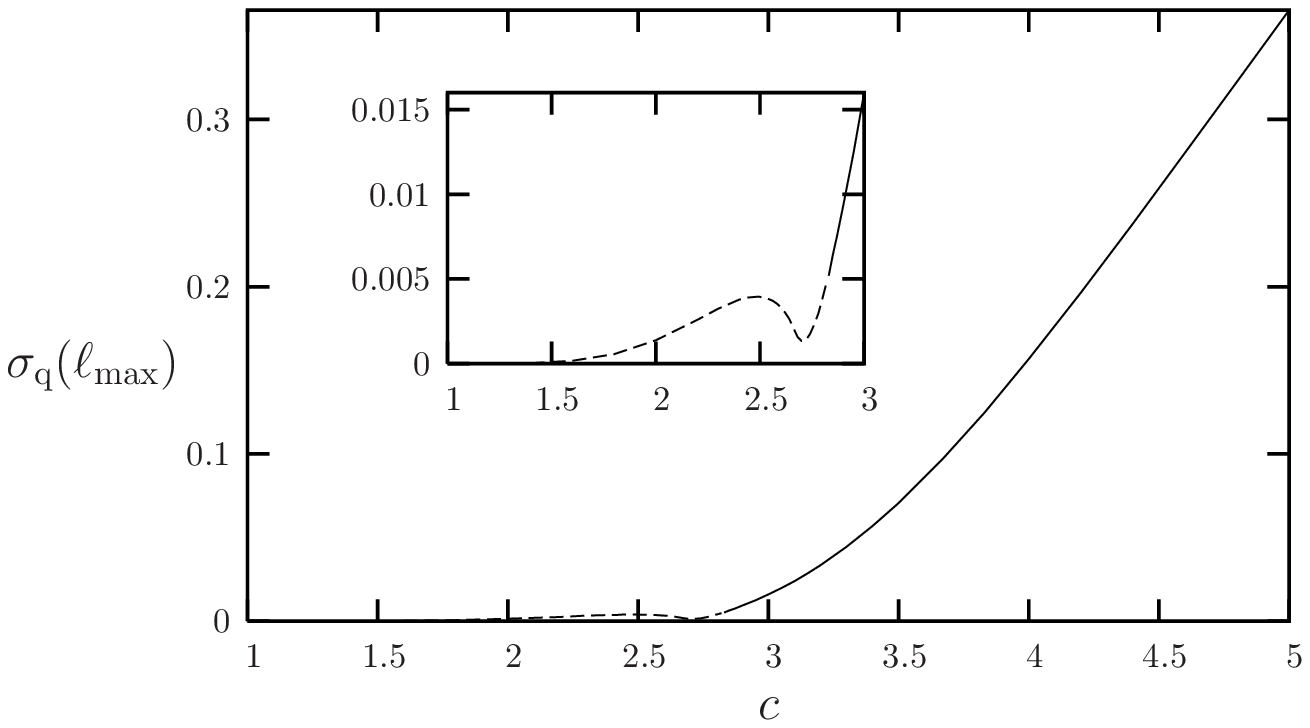}
\end{center}
\caption{The length of longest cycles in Poissonian graphs of mean 
connectivity $c$ (left) and the associated quenched entropy (right). 
Dotted lines in the left panel are the bounds of 
Eq.~(\ref{eq_bounds_poissonian}). The dashed part of the curves corresponds
to the regime $c\le c_{\rm s}^{(+)}$, where we expect the replica
symmetry to be broken. The inset of the right figure shows a blow-up for
small connectivities, the replica symmetric entropy presents a local maximum
(resp. minimum) at $c \approx 2.48$ (resp. $c \approx 2.71$).
}
\label{fig_lmax_poissonian}
\end{figure}

Let us first discuss the computation of $\ell_{\rm max}$.
This quantity can be obtained
from the distribution $v_p$ of the hard fields, independently
of the evanescent ones. Integrating away the evanescent fields in 
Eq.~(\ref{eq_Vpx}), one obtains:
\begin{equation}
v_p = \sum_{k=1}^\infty \tilde{r}_k 
\sum_{p_1,\dots,p_k \in \mathbb{Z}^k} v_{p_1} \dots v_{p_k} 
\delta_{p,e_k(p_1,\dots,p_k)} \ .
\label{eq_vp}
\end{equation}
Besides the population dynamics method, a faster and more precise method
can be devised to solve this equation. Let us for this purpose
define $w_p$, the integrated form of the distribution,
and $\varphi(x)$ the generating function of the $\tilde{r}_k$'s:
\begin{equation}
w_p = \sum_{p'=-\infty}^{p-1} v_p \ , \quad 
\varphi(x) = \sum_{k=1}^\infty \tilde{r}_k x^k \ .
\end{equation}
One can then rewrite Eq.~(\ref{eq_vp}),
after a few lines of computation based on the expression of 
$e_k(p_1,\dots,p_k)$, under the form
\begin{equation}
\begin{cases}
w_{p+1} = 1 - (1-w_p) \varphi'(w_{-p}) \ \\
w_{-p} = 1-(1-w_{p+1})\varphi'(w_{p+1})+ \varphi(w_{-p-1})- \varphi(w_{p+1}) 
\end{cases} \quad {\rm for} \ p \ge 0 \ .
\end{equation}
It is easy to solve them numerically by iteration (both $w_{-p}$ and $1-w_p$ 
vanish exponentially fast when $p \to + \infty$, a cutoff on $p$ can thus
be safely introduced), and to deduce $\ell_{\rm max}$ from the solution $v_p$
(see Eqs.~(\ref{eq_Jkh},\ref{eq_lmax})). We present the results of this
procedure for Poissonian graphs in the left panel of 
Fig.~\ref{fig_lmax_poissonian}, along with the bounds discussed above.

We now sketch the way to compute the expansion of $\ell_{\rm max}$ stated in 
Eq.~(\ref{eq_lmax_expansion}). 
In the limit $\tilde{r}_1 \to 0$, the distribution $v_p$ tends to 
$\delta_{p,0}$. A more precise inspection reveals that 
$v_p = O(\tilde{r}_1^p)$, $v_{-p} = O(\tilde{r}_1^{2p})$ for $p>0$. In order
to obtain Eq.~(\ref{eq_lmax_expansion}), it is thus enough to find
$\{v_{-1}, v_0, v_1, v_2, v_3 \}$ at order $\tilde{r}_1^3$, in function of
the connectivity distribution $\tilde{r}_k$. The result follow by
collecting the terms of order $\tilde{r}_1^3$ in 
Eqs.~(\ref{eq_Jkh},\ref{eq_lmax}). This expansion
could be in principle pursued at any higher order, at the price of more tedious
computations.

If one is not only interested in the length of the longest circuits, but
also in the associated entropy $\sigma_{\rm q}(\ell_{\rm max})$, one has
to solve the complete equation (\ref{eq_Vpx}) on the distribution of both 
hard and evanescent fields. This is easily done by a population dynamics
algorithm, following population of couples $(p,x)$, see the right 
panel of Fig.~\ref{fig_lmax_poissonian} for the results in the Poissonian case.
However, when the value of $\tilde{r}_1$ is too large, there appears an
instability in the resolution of Eq.~(\ref{eq_Vpx}). For the sake of
definiteness let us consider the Poissonian case and postpone a more 
general discussion to the next section. For values of $c$ larger than a 
critical value $c_{\rm s}^{(+)}\approx 2.88$, the evanescent fields 
distributions converge, whereas below $c_{\rm s}^{(+)}$, the iteration brings 
some of them towards 
diverging or vanishing values. The origin of this instability can be traced
back to the behaviour of the original messages $y$ at large but finite $u$.
A closer inspection of numerically obtained histograms of the $y$'s reveals
that in this limit they indeed obey a scaling of the form $y \sim x u^{p-1/2}$,
but $p$ is a relative integer only for $c \ge c_{\rm s}^{(+)}$. For lower
connectivities, a continuously growing fraction of the hard fields are 
half-integers. This fraction reaches one at $c_{\rm s}^{(-)}\approx 2.67$,
below which all $p$'s are half-integers. If one allows the hard fields to
be both integers and half-integers in 
Eqs.~(\ref{eq_Vpx},\ref{eq_Jkh},\ref{eq_Jk}), this instability problem is 
cured, which allowed us to obtain the (dashed) low connectivity part 
of the curves in Fig.~\ref{fig_lmax_poissonian}.
We shall come back in the next section on the 
interpretation of this phenomenon.

\section{Stability of the replica-symmetric ansatz}
\label{sec_stability}

The cavity computations we have presented so far were based on the assumption
of replica symmetry (RS), valid if the space of configurations is smooth 
enough. In disordered systems this assumption can be violated, we shall 
thus investigate its validity in the present model. More
precisely, we consider the local stability of the RS ansatz in the
enlarged space of one step replica symmetry breaking (1RSB) order 
parameters~\cite{MePa_Bethe} (we leave aside the possibility of a
discontinuous transition).
In the 1RSB setting, the messages $y$ are replaced by probability
distributions $Q(y)$ over the states, and the recursion 
$y \leftarrow g_k(y_1,\dots,y_k)$ becomes
\begin{equation}
Q(y) \leftarrow \frac{1}{\cal Z} \int dy_1 Q_1(y_1) \dots dy_k Q_k(y_k) \
\delta (y - g_k(y_1,\dots,y_k) ) \ W(y_1,\dots,y_k)^m \ ,
\label{eq_1rsb}
\end{equation}
where $\cal Z$ is a normalization constant, $m$ is the Parisi 1RSB parameter,
and $W(\{ y_{k \to i} \})$ is a reweighting factor whose
explicit form is not needed here. The distributions $Q$ are themselves
drawn from a distribution over distributions, ${\cal Q}[Q]$.

The replica symmetric solution studied in the main part of the
text is recovered by taking the distributions $Q$ concentrated on a
single value $y$. To investigate its local stability, one gives them
an infinitesimal variance $v$. Expanding Eq.~(\ref{eq_1rsb})
in the limit of vanishing $v$'s, one obtains the following relation:
\begin{equation}
(y,v) \leftarrow \left( g_k(y_1,\dots,y_k), \sum_{j=1}^k \left( 
\frac{\partial g_k(y_1,\dots,y_k)}{\partial y_j} \right)^2 v_j  \right) \ .
\label{eq_rsb_recur}
\end{equation}
For the RS solution to be stable against this perturbation, the variances
of the 1RSB order parameters should decrease upon iterations of the above
relation. This can be studied numerically for any random graph ensemble, 
by iterating the above relation on a population of couples $(y,v)$,
the value of $k$ being drawn from $\tilde{r}_k$. The variances $v$ can
be initially all taken to 1 (note that Eq.~(\ref{eq_rsb_recur}) is linear in 
the $v$'s), in the course of the dynamics the $v$'s are periodically 
divided by a number $\lambda$, chosen each time to maintain the average
value of $v$ constant. After a thermalization phase $\lambda$ converges
(in order to gain numerical precision one computes the average over the 
iterations of $\ln \lambda$), its limit being $>1$ (resp $<1$) if the RS
solution is unstable (resp. stable). This method, pioneered in the context
of the instability of the 1RSB solution in~\cite{MoPaRi}, can be replaced
by the computation of the associated non-linear susceptibility, see for 
instance~\cite{BM}.

For regular random graphs of connectivity $c$, where all RS messages
take the same value $y_{\rm r}(u,c)$ given in Eq.~(\ref{eq_y_regular}),
one can readily compute the value of the stability parameter,
\begin{equation}
\lambda_{\rm r}(u,c) = 
\frac{(2 - u(c-1))^2}{u^2 (c-1)^3} \ .
\label{eq_stab_reg}
\end{equation}
It is easy to check that $0\le \lambda_{\rm r}(u,c)\le (c-1)^{-1} < 1$ 
when $u$ lies in its allowed range $[u_{\rm m}=(c-1)^{-1},\infty[$,
confirming the validity of the RS ansatz. 
It would have been anyhow surprising to discover
an instability in this case where the annealed computation is exact.

Another case which is analytically solvable is the limit $\ell \to 0$ 
(i.e. $u \to u_{\rm m}$). Indeed, we have seen that the messages scales then
as $x (u-u_{\rm m})^2$, and it turns out that
$\partial g_k / \partial y_i \to u_{\rm m}$, independently
of the rescaled messages $x$. Recalling that 
$\sum_{k} \tilde{r}_k k = \tilde{\mu}_1 = u_{\rm m}^{-1}$, one finds
$\lambda = \tilde{\mu}_1^{-1} < 1$ in this limit: for any connectivity 
distribution, the RS ansatz is always stable in the small $\ell$ regime.

All the numerical investigations of $\lambda$ we conducted for ensembles 
with minimal connectivity 3 suggest that in this case 
the replica symmetric solution is stable for all values of $u$. Note that
here the zero temperature limit of $\lambda$ can be studied directly at the
level of evanescent fields, as 
$\partial g_k / \partial y_i \to \partial h_k / \partial x_i$.
We thus conjecture that the whole function $\sigma_{\rm q}(\ell)$ computed
with the RS cavity method is correct for these ensembles, and in particular
the quenched entropy of Hamiltonian circuits $\sigma_{\rm q}(1)$ stated in
Eq.~(\ref{eq_Wormald_refined}).

\begin{center}
\begin{figure}
\includegraphics[width=8cm]{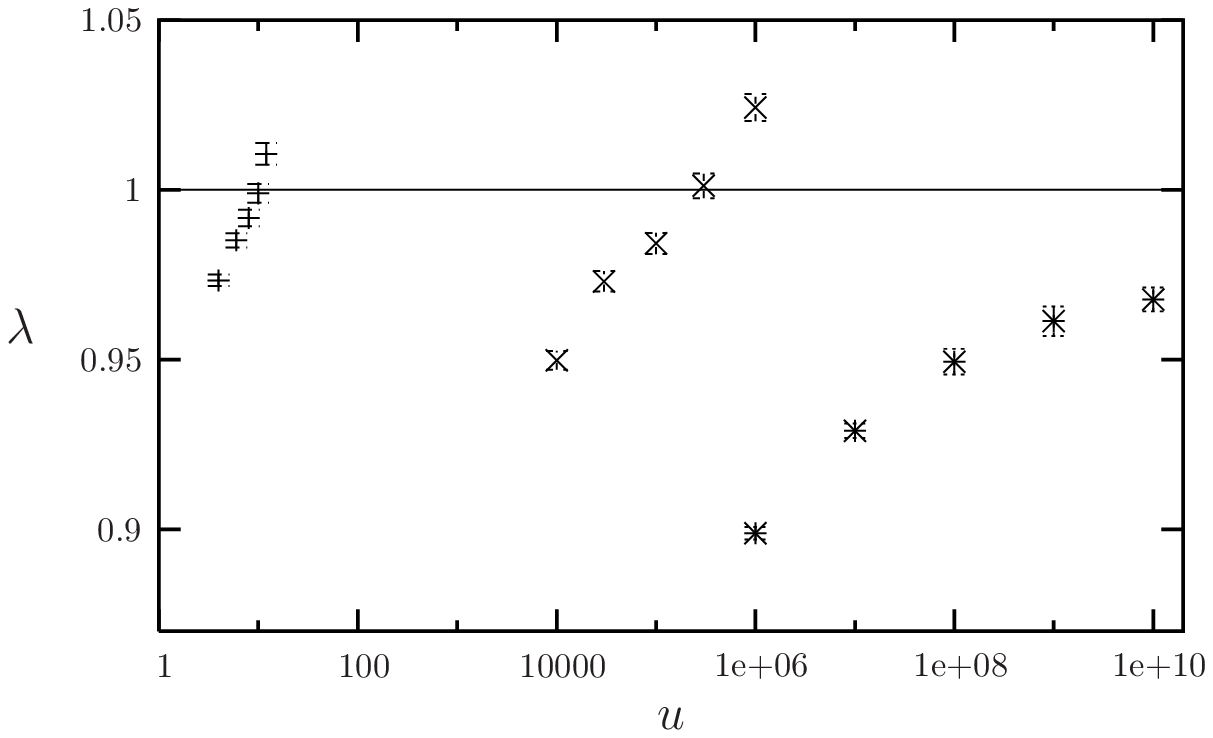} 
\hspace{1cm}
\includegraphics[width=8cm]{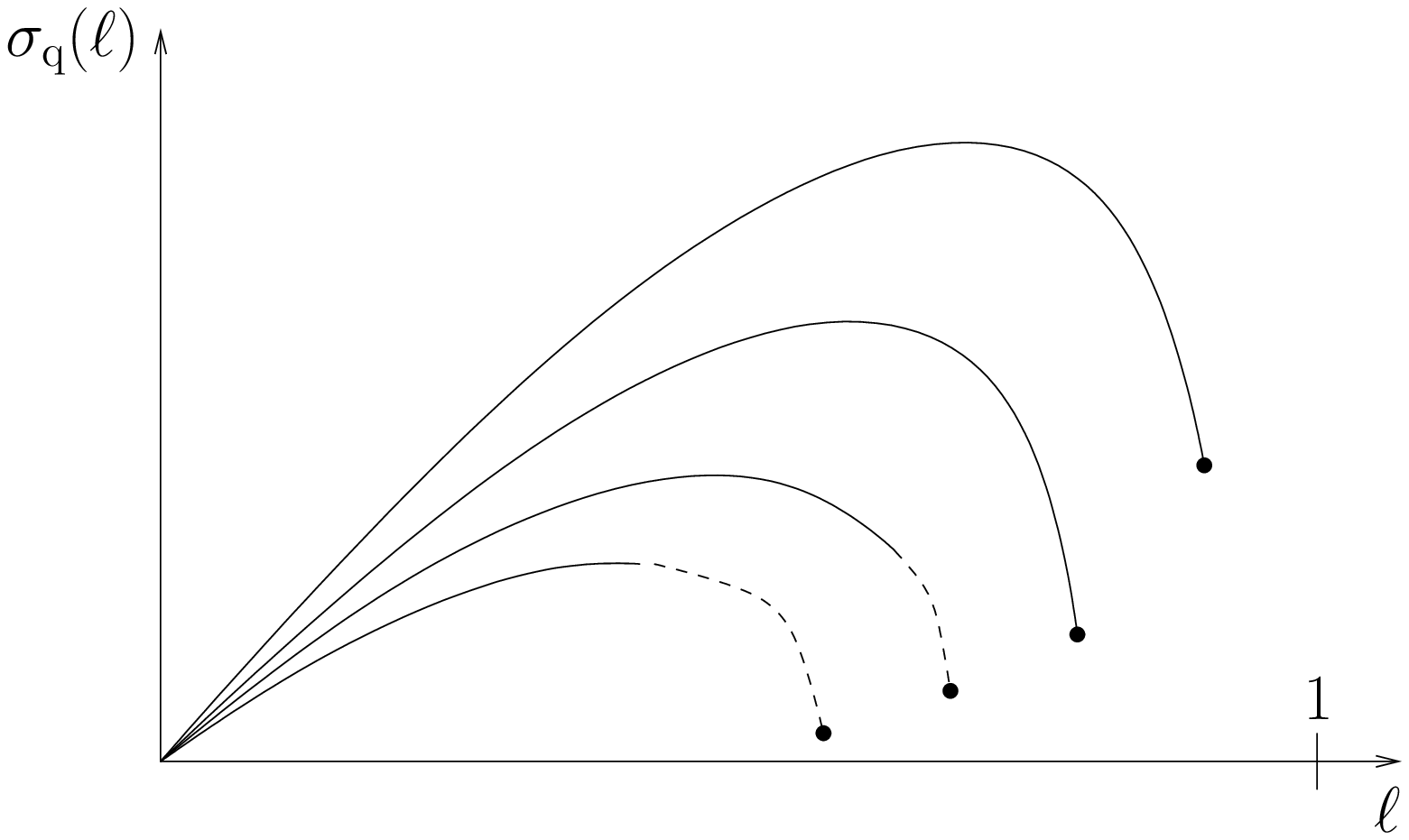} 
\caption{Left: the stability parameter $\lambda$ for Poissonian random
graphs, from left to right $c=1.2$, $2$, $3$. Right:
Sketched behaviour of the quenched entropy for generic families of random
graphs. From top to bottom a control parameter drives the graphs towards 
a continuous percolation transition, the maximal length of the circuits is 
reduced. In the neighborhood of the percolation transition
replica symmetry breaking takes place for large enough circuits, 
and should be taken into account to compute the dashed part of the curve.
}
\label{fig_sketch_sigma}
\end{figure}
\end{center}

The situation is less fortunate for Poissonian graphs. The reader may have
anticipated the appearance of non integer hard fields
in the zero temperature limit for mean connectivities lower than 
$c_{\rm s}^{(+)}$ as an hint of RSB. The datas presented in the left panel
of Fig.~\ref{fig_sketch_sigma} shows indeed that for small $c$,
the stability parameter $\lambda$ crosses 1 when $u$ is increased above some 
finite value $u_{\rm s}(c)$. This critical value of $u$ increases with
the mean connectivity, and an educated guess makes us conjecture that it 
diverges at $c_{\rm s}^{(+)}$. The rightmost curve for $c=3$ shows indeed
$\lambda<1$ for all the values of $u$ we could numerically study. A precise
extrapolation of $u_{\rm s}(c)$ turned out however to be rather difficult.
Note also that the study directly at $u=\infty$ is largely complicated
here by the fact that the hard fields do not take a finite number of distinct
values as is often the case in usual optimization problems~\cite{stab_T0}, 
but extend on the
contrary on all relative integers. In summary, the conjectured scenario is
that at high enough connectivities the whole curve $\sigma_{\rm q}(\ell)$, and
in particular its zero temperature limit, is correctly described by the RS
computation. For lower connectivities there will be a critical length above 
which replica symmetry breaks down.
We also believe that this scenario, sketched in the right part of
Fig.~\ref{fig_sketch_sigma}, is valid not only in the Poissonian case,
but for all families of random graph ensembles (with a fastly decaying
connectivity distribution) with a control parameter which drives the graphs 
towards a continuous percolation transition, the fraction of degree 2 sites 
in the 2-core growing as the transition is approached.

Let us finally propose an interpretation for the occurrence of replica 
symmetry breaking for the largest circuits in presence of a large fraction of 
degree 2 sites in the 2 core, by relating it to an underlying extreme value 
problem~\cite{extreme}. In the discussion of Sec.~\ref{sec_bounds}, one could
indeed tag the edges $l$ of the reduced graph $G_3$ with a strictly positive 
integer, by counting the number of edges of $G_2$ which were collapsed onto 
$l$. The length of a circuit of $G_2$ is thus the weighted length of the
corresponding circuit of $G_3$, i.e. the sum of the labels on the edges it
visits. These weighted lengths are correlated random variables, because
of the structural constraint defining a circuit: for a given graph $G_3$,
not all the sums of $L$ tags correspond to circuits of length $L$. When
the fraction of degree 2 site is small enough, these correlations are
sufficiently weak for the RS ansatz to treat them correctly, when long chains 
of degree 2 vertices become too numerous they somehow pin the longest circuits,
which cluster in the space of configurations and cause the replica symmetry
breaking.

\section{Exhaustive enumerations}
\label{sec_enumerations}

We present in this section the results of the numerical experiments we have 
conducted in order to check our analytical predictions. These experiments
are based on the exhaustive enumeration algorithm of~\cite{Johnson} which 
allows to generate all the circuits of a given graph $G$, and in particular to
compute the numbers ${\cal N}_L(G)$ of circuits of a given length. This 
algorithm runs in a time proportional to the total number of circuits, hence
exponential in the size of the graphs for the cases we are interested in, which
obviously puts a strong limitation on the sizes we have been able to study.

\begin{center}
\begin{figure}
\includegraphics[width=8cm]{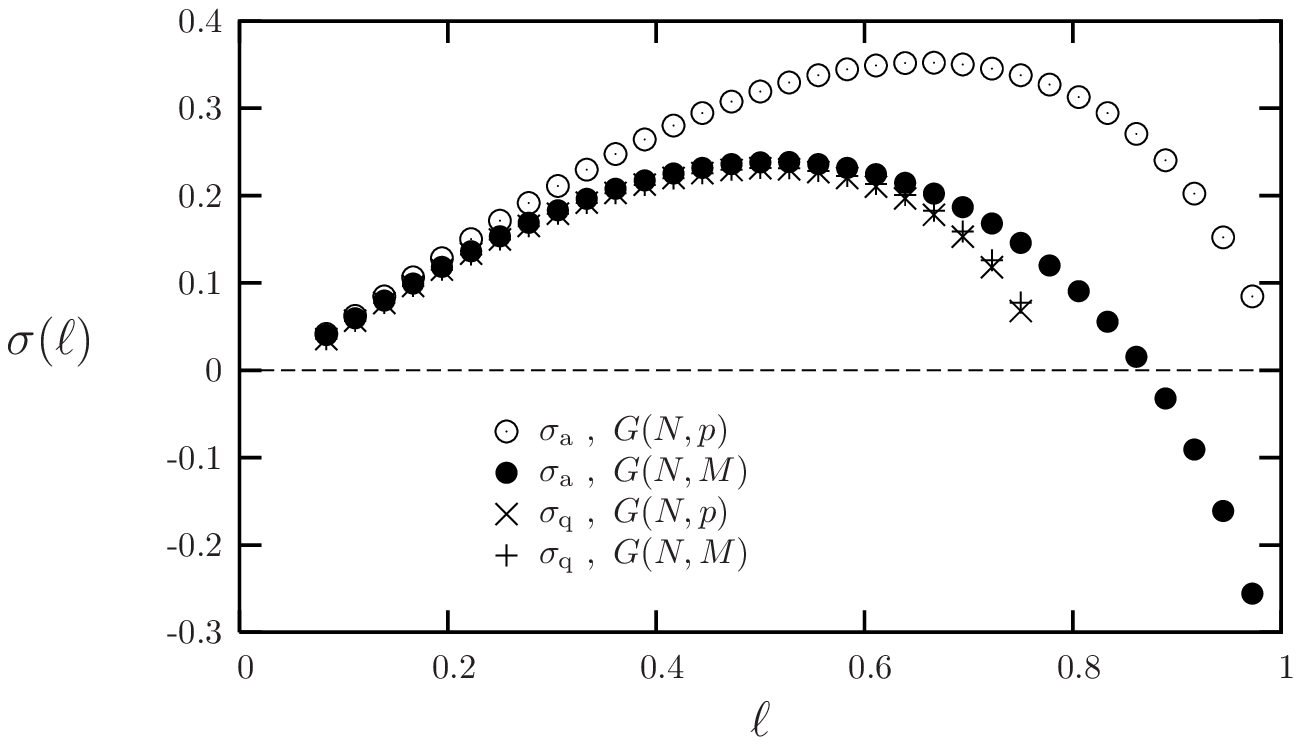} 
\hspace{1cm}
\includegraphics[width=8cm]{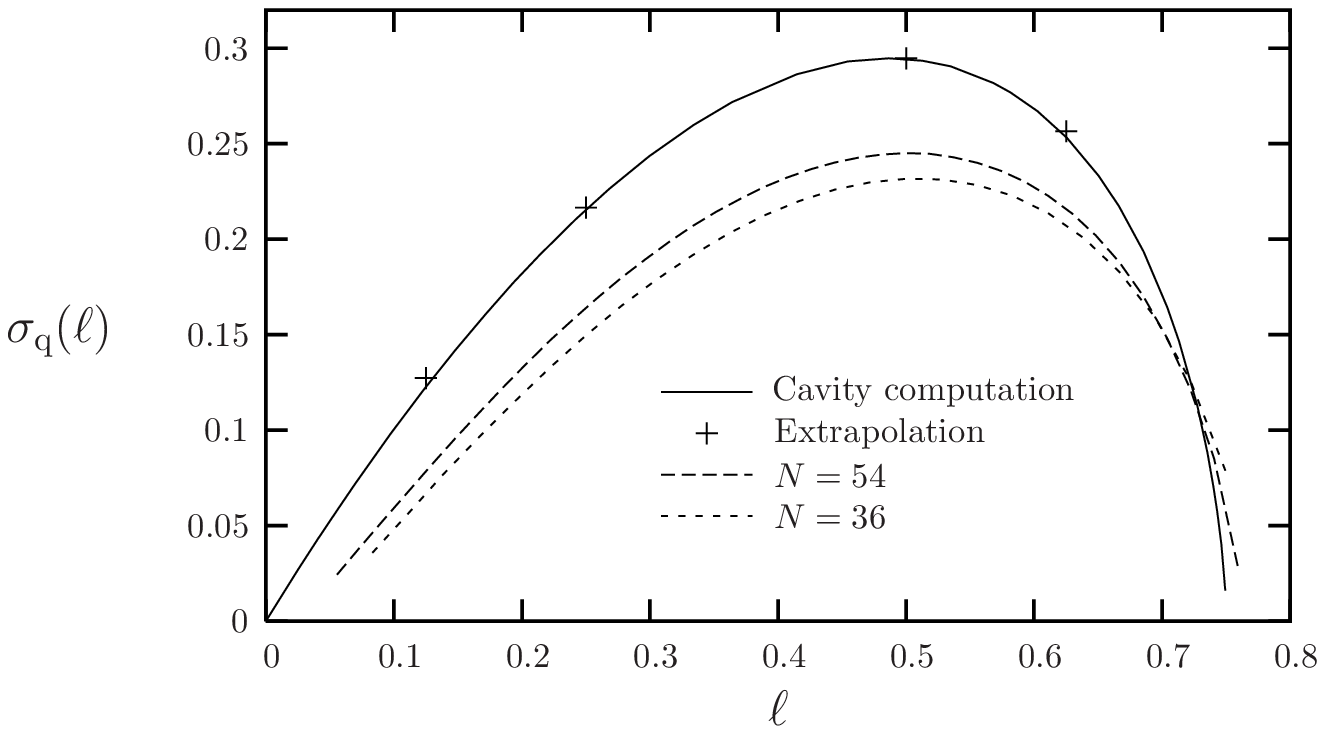} 
\caption{Left: Annealed and quenched entropies for the Erd\"os-R\'enyi 
ensembles $G(N,p)$ and $G(N,M)$ of mean connectivity $c=3$, for graphs of size
$N=36$, computed from the mean and the median of ${\cal N}_L$ on samples of
10000 graphs. Right: the quenched entropy for $G(N,M)$ at
$N=36$, and $N=54$, symbols are the extrapolation in the limit 
$N \to \infty$ from several values of $N$, solid line is the replica symmetric 
cavity computation.
}
\label{fig_sigmas_c3}
\end{figure}
\end{center}

\begin{center}
\begin{figure}
\includegraphics[width=8cm]{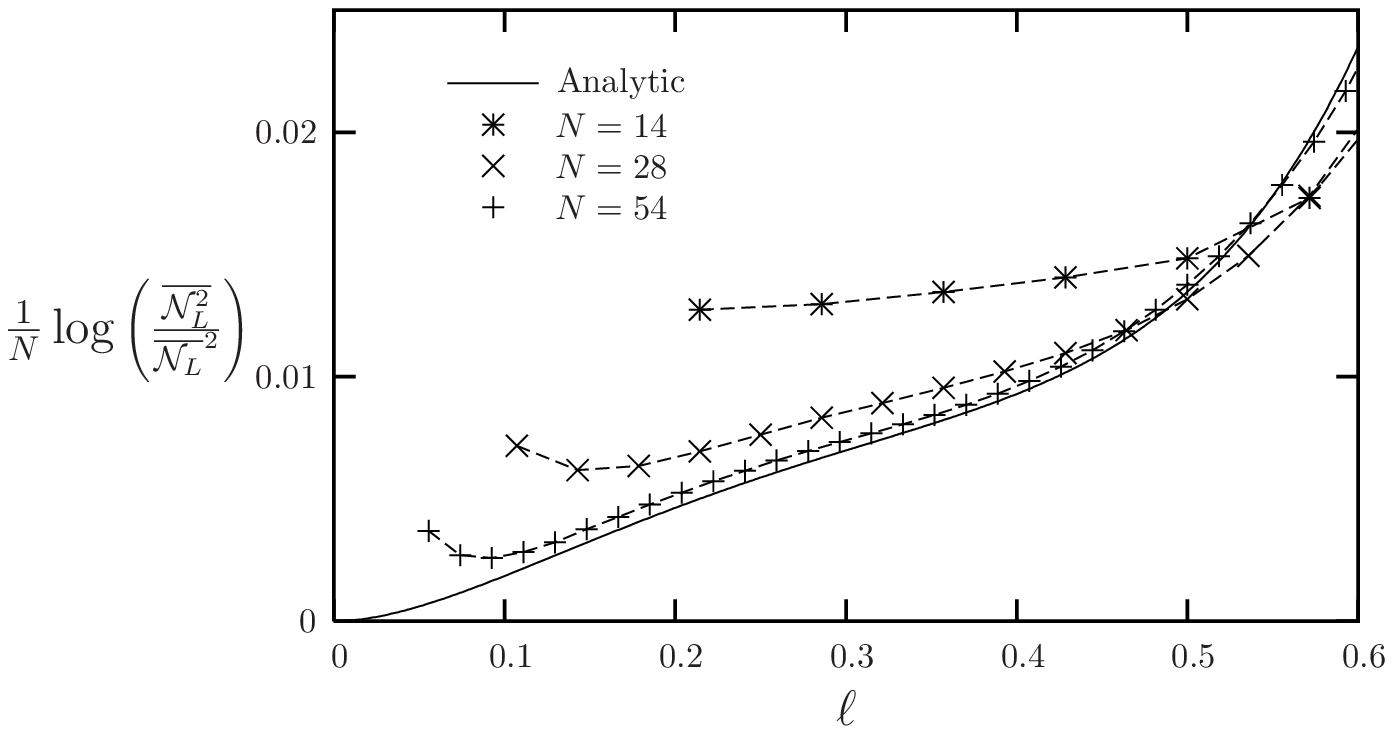} 
\hspace{1cm}
\includegraphics[width=8cm]{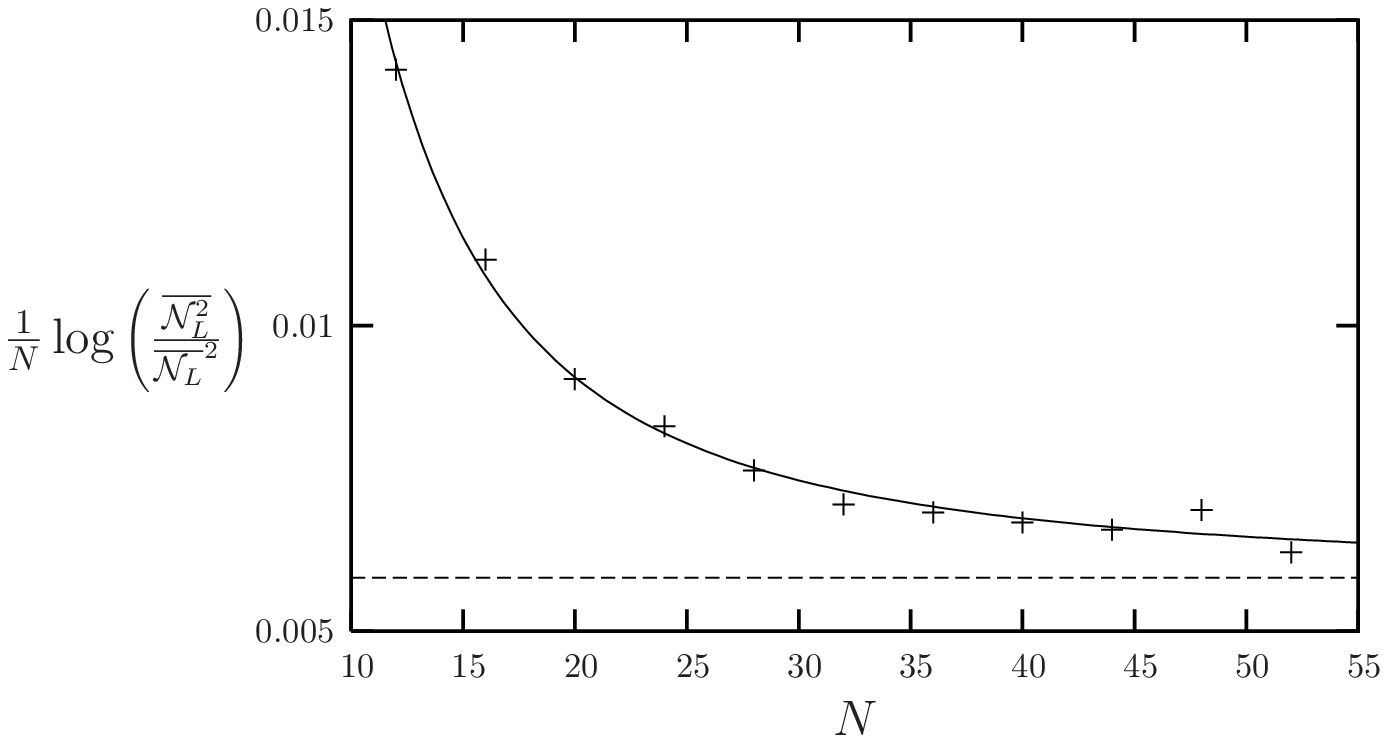} 
\caption{Left: the ratio of the first two moments of ${\cal N}_L$ for $G(N,M)$
at $c=3$, the symbols are numerically determined values which converge in the
large size limit to the solid line, analytically computed in 
App.~\ref{sec_app_combinatorial}. Right: finite size analysis for $\ell =1/4$,
solid line is a best fit of the form $a + b/N + c/N^2$, where $a$ is 
constrained to its analytic value (dashed line), and the form of the fit is
justified in App.~\ref{sec_app_combinatorial}.}
\label{fig_2nd_moment}
\end{figure}
\end{center}

Let us begin with the investigation of the Erd\"os-R\'enyi ensembles
$G(N,p)$ and $G(N,M)$. In the former, each of the $N(N-1)/2$ potential edges
between the $N$ vertices of the graph is present with probability $p$, 
independently of each other, in the latter a set of $M$ among the $N(N-1)/2$ 
edges is chosen uniformly at random. With $p =c/N$ and $M=cN/2$, these two
ensembles are expected to be equivalent in the large-size limit. In particular
the vertex degree distribution converges in both cases to a Poisson law of
mean $c$, the cavity computation thus predicts that their typical properties
should be the same in the thermodynamic limit. This is not true for the
the annealed entropies 
$\sigma_{\rm a}(\ell;N)=\log(\overline{{\cal N}_{\ell N}})/N$ which are easily 
computed exactly even at finite sizes, see App.~\ref{sec_app_combinatorial},
and which remain distinct in the thermodynamic limit. In the left part
of Fig.~\ref{fig_sigmas_c3} we present the annealed and quenched entropies 
for both ensembles, computed from 10000 graphs of size
$N=36$ and mean connectivity $c=3$.
The finite size quenched entropy has been estimated using the median
of the random variables ${\cal N}_L$. The annealed entropies are very different
in both ensembles (and in perfect agreement with the computation of 
App.~\ref{sec_app_combinatorial}), and clearly different from the quenched 
ones. The striking feature of this plot is the almost perfect coincidence of
the median in the two ensembles; this was expected in the thermodynamic limit, 
but is already very clear at this moderate size. On the right panel of
Fig.~\ref{fig_sigmas_c3}, the quenched entropy is plotted for two graph 
sizes, along with its extrapolated values in the thermodynamic limit, which
agrees with the cavity computation.
As argued above, the difference between annealed and quenched entropies can
be also seen in the exponentially larger value of the second moment of
${\cal N}_L$ with respect to the square of the first moment. This fact is 
illustrated in Fig.~\ref{fig_2nd_moment}, where the analytic computation
of the ratio $\log(\overline{{\cal N}_L^2}/ \overline{{\cal N}_L}^2)/N$
presented in App.~\ref{sec_app_combinatorial} is confronted with its
numerical determination.
We also considered the largest circuits in each graph, of length $L_{\rm max}$
and degeneracy ${\cal N}_{L_{\rm max}}$, and computed the averages 
$\overline{L_{\rm max}}/N$ and $\overline{\ln {\cal N}_{L_{\rm max}}}/N$ for
various connectivities. Their extrapolated values in the thermodynamic limit 
are compatible with the predictions $\ell_{\rm max}$ and 
$\sigma_{\rm q}(\ell_{\rm max})$ of the cavity method, within the numerical 
accuracy we could reach. This is true also for connectivities smaller than
$c_{\rm s}^{(+)}$, where we argued above in favor of a violation of the
replica symmetry hypothesis: the corrections due to RSB should be smaller than
the numerical precision we reached.

\begin{center}
\begin{figure}
\includegraphics[width=8cm]{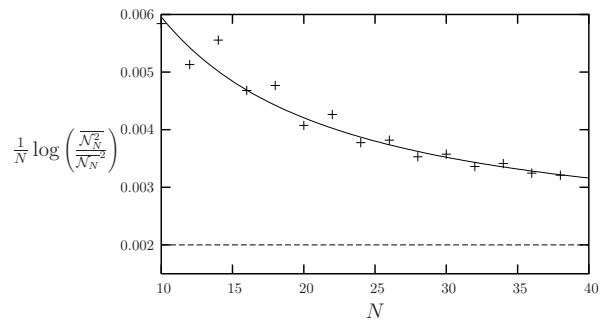} 
\caption{The ratio of the  first two moments of ${\cal N}_N$ for graphs
with an equal fraction of degree 3 and 4 vertices.
Solid line is a best fit of the form $a + b/N + c/N^2$, with $a\approx 0.002$,
as found in App.~\ref{sec_app_combinatorial}.}
\label{fig_2nd_moment_3p4}
\end{figure}
\end{center}

Another set of experiments concerned uniformly generated graphs with an equal
number of degree 3 and 4 vertices. We checked that the probability for such 
graphs to be Hamiltonian converges to 1 when increasing their size. The
values for the annealed and quenched entropies for the Hamiltonian circuits are
too close to be distinguished numerically. However the study of the ratio of
the first two moments of ${\cal N}_N$ (see Fig.~\ref{fig_2nd_moment_3p4})
indicates that they should be strictly distinct in the thermodynamic limit.

\section{Conclusions and perspectives}
\label{sec_conclusion}

Let us summarize the main results presented in this paper. We have proposed
an approximative counting algorithm that runs in a linear time with respects
to the size of the graph. We also presented an heuristic method to compute
the typical number of circuits in random graph ensembles, which yields 
a quantitative refinement of Wormald's conjecture on the typical number of
Hamiltonian cycles in ensembles with minimal degree 3 
(Eq.~(\ref{eq_Wormald_refined})) and a new 
conjecture on the maximal length of circuits in ensembles with a small
fraction of degree 2 vertices in their 2 cores (Eq.~(\ref{eq_lmax_expansion})).

Several directions are opened for future work. First of all we believe
that a rigorous proof of Wormald's conjecture, which seems difficult to reach
by variations around the second moment method, could be obtained by statistical
mechanics inspired techniques. In recent years there has been indeed a series
of mathematical achievements in the formalization of the kind of method used
in this article. One line of research is based on Guerra's interpolation 
method~\cite{Guerra}, and culminated in Talagrand's proof of the correctness 
of the Parisi free-energy formula for the Sherrington-Kirkpatrick 
model~\cite{Talagrand}. These ideas have 
also been applied to sparse random graphs in~\cite{FrLe,FrLeTo}. 
Alternatively the local weak convergence method of Aldous~\cite{Aldous} 
has been successfully applied to similar counting problems in 
random graphs~\cite{Counting}.

There has also been a recent interest~\cite{MoRi,PaSl,ChCh} in the 
corrections to the Bethe approximation for general graphical models. It 
would be of great interest to implement these refined approximations for the
counting problem considered in this paper. This should lead on one hand to 
a more precise counting algorithm, and on the other hand give access to the 
finite-size corrections of the quenched entropy. We expect in particular
that the difference between circuits and unions of vertex disjoint circuits 
will become relevant for these corrections.

The convergence in probability of $\log {\cal N}_L /N$ expressed by
Eq.~(\ref{eq_cvp}) can a priori be promoted to a stronger large deviation
principle: according to the common wisdom, the finite deviations of this
quantity from $\sigma_{\rm q}$ are exponentially small. A general method
for computing these rate functions has been presented in~\cite{Olivier} and
could be of use in the present context. An interesting question could be
to compute the exponentially small probability that a random graph is not
Hamiltonian in ensembles where typical graphs are so.

In the algorithmic perspective, one could try to take advantage of the
local informations provided by the messages. In particular they could be useful
to explicitly construct long cycles, in a ``belief inspired decimation'' 
fashion~\cite{SP}: most probable edges in the current probability law would
be recursively forced to be present, and the BP equations re-runned in the new
simplified model.

The neighborhood of the percolation transition should also be investigated
more carefully, in particular the effects of replica-symmetry breaking
onto the structure of the configuration space.

The case of heavy-tailed (scale-free) degree distributions deserves also
further work. The assumption of fast decay we made here is indeed crucial
for some of our results: Bianconi and Marsili showed in~\cite{BiMa}
that scale-free graphs, even with a minimal connectivity of 3, can fail
to have Hamiltonian cycles. Other random graph models (generated by a growing
process~\cite{growing}, or incorporating correlations between vertex 
degrees~\cite{BiMa2}) could also been investigated.

Let us finally mention two closely related problems which are currently 
studied with very similar means. Circuits can be defined as a particular case
of $k$-regular graphs, with $k=2$. Replacing the number of allowed edges
around any site from 2 to $k$ in Eq.~(\ref{eq_wi}), one can similarly
study the number of $k$-regular subgraphs in random graph ensemble.
The case $k=1$ corresponds to matchings, which was
largely studied in the mathematical
literature~\cite{matching,matching2} and have been reconsidered by statistical
mechanics methods in~\cite{Lenka}. The appearance of $k \ge 3$-regular
subgraphs in random graphs was first considered 
in~\cite{Bollobas_k}, see~\cite{Martin} for a statistical mechanics treatment. 

\acknowledgments

We warmly thanks R\'emi Monasson with whom the first steps of this work have
been taken.
We also acknowledge very useful discussions with 
Ginestra Bianconi,
Andrea Montanari,
Andrea Pagnani,
Federico Ricci-Tersenghi,
Olivier Rivoire,
Martin Weigt and
Lenka Zdeborov\'a.

The work was supported by EVERGROW, integrated project No. 1935 in the
complex systems initiative of the Future and Emerging Technologies directorate 
of the IST Priority, EU Sixth Framework.

\appendix

\section{Combinatorial approach}
\label{sec_app_combinatorial}

\subsection{Notation and definitions}
We collect in this appendix the combinatorial arguments for the
computation of the
first and second moment of the number of circuits in various random graph
ensembles. Let us denote ${\cal N}_L(G)$ the number of circuits of length
$L$ in a graph $G$, and ${\cal C}_L$ the set of circuits of length $L$ in
the complete graph of $N$ vertices, its cardinality being
\begin{equation}
|{\cal C}_L|= {\cal M}_L = \frac{1}{2L} \frac{N!}{(N-L)!} \ .
\end{equation}
Indeed, choosing such a circuit amounts to select an ordered list of 
the $L$ vertices it will visit, modulo the orientation and the starting point
of the tour. Introducing $\mathbb{I}(H;G)$ the indicator function equal to
$1$ if $H$ is a subgraph of $G$, $0$ otherwise, we can write
\begin{equation}
{\cal N}_L(G) = \sum_{C \in {\cal C}_L} \mathbb{I}(C;G) \ .
\label{combin_def_N}
\end{equation}
Let us now describe the random graph ensembles we shall consider in the 
following. The first two are the classical Erd\"os-R\'enyi random graph 
ensembles.
In $G(N,p)$, each of the $N(N-1)/2$ edges is present with
probability $p$, independently of the others. In $G(N,M)$, a set of 
$M$ distinct edges is chosen uniformly at random among the $N(N-1)/2$ possible 
ones. We shall concentrate on the thermodynamic limit
$N \to \infty$, $p=c/N$ and $M=cN/2$ with the mean connectivity $c$ 
kept finite. In this regime $G(N,p)$ and $G(N,M)$ are essentially
equivalent: drawing at random from $G(N,p)$ amounts to draw $M$ from
a binomial distribution of parameters $(p,N(N-1)/2)$, and then drawing
at random a graph from $G(N,M)$. In the limit described above,
the number of edges in $G(N,p)$ is weakly fluctuating around $M=cN/2$.
Moreover the degree of a given vertex in the graph converges in 
both cases to a Poisson random variable of parameter $c$.

For an arbitrary degree distribution $q_k$ of mean $c$, 
one can define the uniform 
ensemble of graphs obeying this constraint of degree distribution. A practical
way of drawing a graph from this ensemble is the so-called configuration 
model~\cite{conf_Bollo}, defined as follows. Each of the vertices is randomly 
attributed a degree, in such a way that $N q_k$ vertices have degree
$k$ (we obviously skip some technical details~\cite{MR}: $q_k$ should be
a function of $N$, such that $N q_k$ is an integer). 
$2k$ half-links goes out of each vertex of degree $k$. 
Then one generates a random matching of the $cN=2M$ half-links and puts an edge
between sites which are matched. In general one obtains in this way a 
multigraph, i.e. there appear edges linking one vertex with itself, or 
multiple edges between the same pair of vertices. However, discarding the
non-simple graphs leads to an uniform distribution over the simple 
ones~\cite{Wormald_review}. To compute averages over the graph
ensemble, one can thus use the configuration model and condition on the
multigraph to be simple. For clarity we shall denote ${\cal N}_L^*(G)$ the
number of circuits in the unconditioned multigraph ensemble.
Note also that regular random graphs are a
particular case of this ensemble, with $q_k = \delta_{k,c}$.

\subsection{First moment computations}
\subsubsection{Generalities}
Taking the average over the graphs of Eq.~(\ref{combin_def_N}) leads to
\begin{equation}
\overline{{\cal N}_L(G)} = \sum_{C \in {\cal C}_L} \overline{\mathbb{I}(C;G)}
= {\cal M}_L  {\cal P}_L 
\end{equation}
for the ensembles we are considering, where the probability 
${\cal P}_L = \overline{\mathbb{I}(C;G)}$ for a circuit $C \in {\cal C}_L$ to
be present is independent of $C$. Before inspecting the various cases, 
let us state the asymptotic behaviour
of ${\cal M}_L$ in the limit $N,L \to \infty$, $\ell = L/N$ finite, obtained
with the Stirling formula:
\begin{eqnarray}
{\cal M}_{\ell N} &=& 
\frac{1}{N} \frac{1}{2 \ell \sqrt{1-\ell}} N^L
e^{N(-h(1-\ell) - \ell)} (1+O(N^{-1}))  \quad \mbox{for} \ 0<\ell<1 \ , \\
{\cal M}_N &=& \frac{1}{\sqrt{N}} \sqrt{\frac{\pi}{2}} 
N^N e^{-N} (1+O(N^{-1})) \ .
\end{eqnarray}
In the first formula we have introduced the function $h(x)=x \ln x$.

\subsubsection{Erdos-Renyi ensembles}
In $G(N,p)$ the probability
${\cal P}_L$ has a very simple expression, ${\cal P}_L=(c/N)^L$.
The mean number of circuits thus reads
\begin{equation}
\overline{{\cal N}_L(G)} = 
\frac{1}{2L} \frac{N!}{(N-L)!} \left( \frac{c}{N} \right)^L
= \frac{1}{N} \frac{1}{2\ell \sqrt{1-\ell}} e^{N \sigma(\ell)}
(1+O(N^{-1})) \ ,
\label{eq_ER1_ann}
\end{equation}
where the first expression is valid for any $N,L$, and the second one has
been obtained in the thermodynamic limit with $0<\ell<1$ 
The annealed entropy for this first ensemble is:
\begin{equation}
\sigma(\ell) = -(1-\ell)\ln(1-\ell) + \ell ((\ln c)-1) \ .
\label{eq_ER1_sigma}
\end{equation}
Note that if $\ell=1$, the algebraic prefactor in (\ref{eq_ER1_ann}) is
slightly different,
\begin{equation}
\overline{{\cal N}_N(G)} = \frac{1}{\sqrt{N}} \sqrt{\frac{\pi}{2}} 
e^{N((\ln c)-1)} (1+O(N^{-1})) \ .
\end{equation}

In $G(N,M)$ the probability ${\cal P}_L$ reads
\begin{equation}
{\cal P}_L = \frac{\left( {N \choose 2} -L \right)! }
{\left( {N \choose 2}\right)!} \frac{M!}{(M-L)!} \ .
\end{equation}
Obviously this expression has a meaning only for $L \le M$, as there cannot
be circuits longer than the total number of edges.
This gives an exact expression for 
$\overline{{\cal N}_L(G)} = {\cal M}_L {\cal P}_L$ for any $N$ and 
$L \le \min(N,M)$. 
The expansion in the thermodynamic limit with $0<\ell<\min(1,c/2)$
leads to
\begin{equation}
\overline{{\cal N}_L(G)} = \frac{1}{N} 
\frac{e^{\ell(\ell+1)}}{2 \ell \sqrt{1-\ell}\sqrt{1-\frac{2\ell}{c}}}
e^{N \sigma(\ell)} (1+O(N^{-1})) \ .
\label{eq_ER2_ann}
\end{equation}
with the annealed entropy
\begin{equation}
\sigma(\ell) = -(1-\ell) \ln (1-\ell) + 
\left(\ell-\frac{c}{2}\right) \ln \left(1 - \frac{2 \ell}{c} \right) +
\ell ((\ln c)-2) \ .
\label{eq_ER2_sigma}
\end{equation}
Again the different algebraic prefactor in (\ref{eq_ER2_ann}) can be 
easily computed also for $\ell = \min(1,c/2)$.

Let us now make a few comments on these results. First, when $c<1$,
both annealed entropies are negative for all values of $\ell>0$ where they
are well defined. Consequently $\overline{{\cal N}_L(G)}$ is exponentially 
small in the
thermodynamic limit, and thanks to the so-called Markov inequality (or
first moment method) valid for positive integer random variables,
\begin{equation}
{\rm Prob} [{\cal N}_L(G) >0 ] \le \overline{{\cal N}_L(G)} \ ,
\end{equation}
with high probability there are no circuits of extensive lengths in these
graphs. This could be expected: the percolation transition occurs at
$c=1$, in this non percolated regime the size of the largest component is of 
order $\ln N$, and thus extensive circuits cannot be present.

As a second remark, let us note that
for $c>1$, the annealed entropy of the first ensemble
is strictly positive for $\ell \in ]0,\ell_{\rm a}(c)[$,
with $\ell_{\rm a}(c)$ an increasing function which reaches the value $1$ in
$c=e$. The average number of such circuits is in consequence exponentially 
large. However one can easily convince oneself that this cannot be the typical 
behaviour. Indeed, it turns out that $\ell_{\rm a}(c)$ is larger than the
typical number of vertices in the 2-core of the graphs 
$\ell_{\rm core}(c)$. When $\ell$ belongs to the interval
$]\ell_{\rm core}(c),\ell_{\rm a}(c)[$,
typically the graphs cannot contain circuits of $L=\ell N$ edges, however
an exponentially small fraction of the graphs have 2-core larger than their
typical sizes. These untypical graphs contribute with an exponential
number of circuits to the annealed mean $\overline{{\cal N}_L(G)}$, which
is in consequence not representative of the typical behaviour of the ensemble.

Finally, let us underline that the annealed entropies 
(\ref{eq_ER1_sigma},\ref{eq_ER1_sigma}) for the two ensembles
are definitely different. For instance, in the second ensemble, the entropy is
defined only for $\ell \le c/2$: the number of edges in the graph being
fixed at $M = N c/2$, no circuits can be longer than the number of edges.
On the contrary, in the first ensemble, arbitrary large deviations of the
number of edges from its typical value are possible, even if with an 
exponential small probability.

\subsubsection{Arbitrary connectivity distribution and regular}
The expectation of the number of circuits of length $L$
in the multigraph ensemble extracted with the configuration model was
presented in~\cite{BiMa}. For the sake of completeness and to make the
study of the second moment simpler we reproduce the argument here.
In this case one has
\begin{equation}
{\cal P}_L = \frac{1}{{N \choose L}} \left( \underset{\{L_k \}}{\sum}
\prod_{k=2}^\infty {N q_k\choose L_k} (k(k-1))^{L_k} \right) 
\frac{(cN-2L-1)!!}{(cN-1)!!} \ ,
\label{eq_combin_PL_arb}
\end{equation}
where the sum is over $L_2,L_3,\dots$ positive integers constrained by
$\sum_{k=2}^\infty L_k=L$, and we used the classical notation 
$(2p-1)!!=(2p-1)(2p-3)\dots 1$. The $L_k$'s are the number of sites of
degree $k$ in the circuit, which are to be distributed among the $N q_k$
sites of degree $k$. The term $(k(k-1))^{L_k}$ accounts for the choice of
the half links around each site, and finally the ratio of the double
factorials is the probability that the matching of half-links contains the
desired configuration. Introducing the integral representation of the
Kronecker symbol, $\delta_n=\oint (d\theta/2 i \pi) \theta^{-n-1}$, where
$\theta$ is a complex variable integrated along a closed path around the
origin, this expression can be simplified in
\begin{equation}
{\cal P}_L = \frac{1}{{N \choose L}} \frac{(cN-2L-1)!!}{(cN-1)!!}
\oint \frac{d\theta}{2i\pi} \theta^{-L-1}
\prod_{k=2}^\infty (1+ \theta k(k-1))^{N q_k} \ .
\end{equation}
In the thermodynamic limit the integral can be evaluated by the saddle-point
method, combining the expansion with the one of ${\cal M}_L$ yields
\begin{eqnarray}
\overline{{\cal N}_L^*(G)} &\doteq& e^{N \sigma(\ell)} 
\ ,\\
\sigma(\ell) &=& \frac{1}{2} h(c-2\ell) - \frac{1}{2} h(c) + h(\ell)
+ \underset{\theta}{{\rm ext}} \left[ 
\sum_{k=2}^\infty q_k \ln(1+k(k-1)\theta) - \ell \ln \theta \right] \ , 
\label{eq_sigmaa}
\end{eqnarray}
where here and in the following $\doteq$ stands for equivalence upto
subexponential terms, i.e. $x_N \doteq y_N$ means $(1/N)\log(x_N/y_N)\to 0$
as $N \to \infty$.

In the regular case one has
\begin{equation}
{\cal P}_L = (c(c-1))^L \frac{(cN-2L-1)!!}{(cN-1)!!} \ ,
\end{equation}
from which the prefactors are more easily computed
\begin{eqnarray}
\overline{{\cal N}_L^*(G)} &=& \frac{1}{N} \frac{1}{2\ell \sqrt{1-\ell}}
e^{N\sigma(\ell)}(1+O(N^{-1})) \quad {\rm for} \ 0<\ell<1 \ , \\
\overline{{\cal N}_N^*(G)} &=& \frac{1}{\sqrt{N}} \sqrt{\frac{\pi}{2}} 
e^{N\sigma(1)}(1+O(N^{-1})) \ , \\
\sigma(\ell) &=& -h(1-\ell) - \frac{1}{2} h(c)+ \frac{1}{2} h(c-2\ell) + 
\ell \ln(c(c-1)) \ .
\end{eqnarray}
Moreover the conditioning on the multigraph being simple can be explicitly
done in the regular case~\cite{Garmo}, thanks to the relative concentration
of ${\cal N}_L^*(G)$. This yields
\begin{equation}
\frac{\overline{{\cal N}_L(G)}}{\overline{{\cal N}^*_L(G)}}
\to \exp\left[ 
\frac{(c-2)\ell}{c-2\ell} \left( 1+ \frac{c(1-\ell)}{c-2\ell}\right)
\right] \quad {\rm for} \ 0<\ell \le 1 \ .
\end{equation}
We checked that the numerical findings of~\cite{MaMo} were in perfect
agreement with these exact results.

Note that in the regular case, this conditioning modifies the value
of $\overline{{\cal N}^*_L(G)}$ only by a constant factor, thus the annealed 
entropy is the same in the graph and in the multigraph ensemble. It is not
clear to us whether this fact should remain true for arbitrary connectivity
distributions.

\subsection{Second moment computations}
\subsubsection{Generalities}
\begin{figure}
\includegraphics[width=8cm]{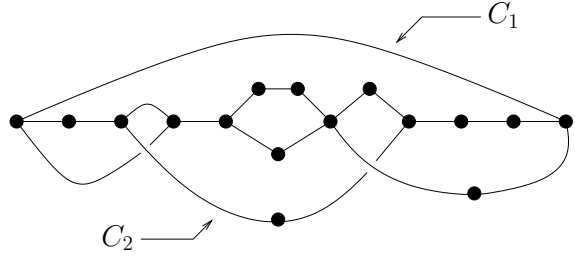}
\caption{The union of two circuits. The vertices of $C_1$ are on and above the
horizontal central line, those of $C_2$ on and below. In this drawing $L=13$, 
$X=3$, $Y=6$, $Z=1$, $n_1=n_2=n_3=1$.}
\label{fig_union}
\end{figure}

We now turn to the computation of the second moment of the number of circuits,
which has been inspired by~\cite{Garmo}. 
Taking the square of Eq.~(\ref{combin_def_N}) and averaging over the
ensemble leads to
\begin{eqnarray}
\overline{{\cal N}_L^2(G)} = \sum_{C_1,C_2 \in {\cal C}_L} 
\overline{\mathbb{I}(C_1;G)\mathbb{I}(C_2;G) } &=& \overline{{\cal N}_L(G)}
+ \sum_{C_1 \neq C_2 \in {\cal C}_L} 
\overline{\mathbb{I}(C_1;G)\mathbb{I}(C_2;G) }  \\
&=& \overline{{\cal N}_L(G)} +
\sum_{X,Y,Z} {\cal M}_{LXYZ} {\cal P}_{LXYZ} \ .
\label{eq_combin_2nd}
\end{eqnarray}
We have indeed isolated the term $C_1 = C_2$ in the sum, which is readily 
computed, from the off-diagonal terms. The last expression is more easily 
understood after having a look at Fig.~\ref{fig_union}, where we sketched
the shape of the union of two distinct circuits. This pattern
is characterized by $X$, the number of common paths shared by $C_1$ and $C_2$,
$Y$, the number of edges in these paths, and $Z$, the number of vertices
which belongs to both circuits but are not neighbored by any common edge.
One finds $2 X$ vertices at the extremities of the common paths, $Y-X$
vertices in the interior of the common paths, hence $X+Y+Z$ vertices belong
to both circuits, $N-2L + X+Y+Z$ to none of them. In consequence the sum
is over $X,Y,Z$ non-negative integers subject to the constraints:
\begin{equation}
2 L - N \le X+Y+Z \le L \ , \quad X \le Y \ , \quad X=0 \Rightarrow Y=0 \ .
\end{equation}
${\cal M}_{LXYZ}$ is the number of pairs of distinct circuits of the complete
graph whose union has the characteristics $(X,Y,Z)$, and ${\cal P}_{LXYZ}$ is
the (ensemble-dependent) probability that such pattern appears in a random
graph.
Let us show that
\begin{equation}
{\cal M}_{LXYZ} =\frac{1}{4} 2^X 
\frac{N! \ ((L-Y-1)!)^2}{X! \ Z! \ ((L-X-Y-Z)!)^2 \ (N-2L+X+Y+Z)!} \
{Y-1 \choose X-1} \ ,
\label{eq_combin_MLXYZ}
\end{equation}
where the combinatorial factor ${-1 \choose -1}$ is by convention set to $1$.
To construct such a pattern, one has to choose among the $N$ vertices those
which are in $C_1$ but not in $C_2$, in $C_2$ but not in $C_1$ (both of these
categories contain $L-X-Y-Z$ vertices), those in the common paths ($X+Y$) and
those shared by the circuits but with no adjacent common edges ($Z$). This
can be done in
\begin{equation}
\frac{N!}{Z! \ (X+Y)! \ ((L-X-Y-Z)!)^2 \ (N-2L+X+Y+Z)!}
\end{equation}
distinct ways.

Let us call $n_i$ the number of common paths of $i$ edges, for 
$1 \le i \le L-1$, which obey the constraints $\sum_i n_i = X$ and
$\sum_i i n_i = Y$. The $X+Y$ sites can be distributed into such 
an unordered set of unorientated paths in
\begin{equation}
\frac{(X+Y)!}{2^X} \frac{1}{\prod_{i=1}^{L-1} n_i!}
\end{equation}
distinct ways. We have to sum this expression on the values of $n_i$ satisfying
the above constraints. 
By picking up the coefficient of $t^Y$ in
\begin{equation}
(t+t^2+\dots+t^{L-1})^X = \sum_{n_1,\dots,n_{L-1}} 
\frac{X!}{\prod_{i=1}^{L-1} n_i!} t^{\sum_i i n_i} \delta_{X,\sum_i n_i}
= t^X \left(\frac{1-t^{L-1}}{1-t}\right)^X \ ,
\end{equation}
one finds that
\begin{equation}
\sum_{n_1,\dots,n_{L-1}} \frac{1}{\prod_{i=1}^{L-1} n_i!}
\delta_{X,\sum_i n_i} \delta_{Y,\sum_i i n_i}
= \frac{1}{X!} {Y-1 \choose X-1} \ .
\end{equation}
When $X=Y=0$ this factor should be one, in agreement with the above convention.

Finally $C_1$ is formed by choosing an ordered list of the $L-X-Y-Z$ vertices
which belongs only to it, the $Z$ isolated common vertices, and of the $X$ 
orientated common paths, modulo the starting point and the global orientation
of this tour, hence a factor
\begin{equation}
\frac{(L-Y-1)!}{2} 2^X \ , 
\end{equation}
and the same arises when constructing $C_2$. Eq.~(\ref{eq_combin_MLXYZ}) is
obtained by multiplying the various contributions.

In the thermodynamic limit with $(x,y,z)=(X/N,Y/N,Z/N)$ kept finite,
Stirling formula yields
\begin{eqnarray}
{\cal M}_{LXYZ} &\doteq& 
N^{N(2\ell -y)} \exp[N m(l,x,y,z)]  \ , \\
m(l,x,y,z)&=& y - 2 \ell + x\ln 2 - 2 h(x) + h(y) - h(z) - h(y-x) \nonumber \\
&+& 2 h(\ell -y) - 2 h(l-x-y-z) - h(1- 2 \ell + x+y+z) \ .
\label{eq_combin_m2}
\end{eqnarray}

\subsubsection{Erdos-Renyi ensembles}
For both $G(N,p)$ and $G(N,M)$ ensembles, the probability 
${\cal P}_{LXYZ}$ depends only on the number of edges present in the 
union of the two circuits, ${\cal P}_{LXYZ}={\cal P}_{2L-Y}$.
For the non trivial range of parameters $\ell,c$ where the first moment is
exponentially large, the first term in Eq.~(\ref{eq_combin_2nd}) can be
neglected. The sum over $(X,Y,Z)$ can be evaluated with the saddle-point
method, yielding
\begin{equation}
\overline{{\cal N}_L^2(G)} \doteq \exp[N \tau(\ell) ] \ , \quad
\tau(\ell) = \underset{y}{\max} [p(2 \ell -y) + \widehat{m}(\ell,y)] \ ,
\end{equation} where 
\begin{equation}
p(\ell) =
\begin{cases}
\ell \ln c & {\rm for} \ G(N,p)  \\
\frac{1}{2} h(c) - \frac{1}{2} h(c -2 \ell) - \ell & {\rm for} \ G(N,M)
\end{cases} \ ,
\end{equation}
 and we introduced
\begin{eqnarray}
\widehat{m}(\ell,y) &=& \underset{x,z}{\max} \ m(\ell,x,y,z) \label{eq_m1} \\
&=& - 2 h(1-\ell) + y - 2 \ell + h(y)  \nonumber \\
& & + \underset{x}{\max}\   \left[ x\ln 2 - 2 h(x) +h(1-x-y) - h(y-x) 
 + 2 h(\ell -y) - 2 h(l-x-y) \right] \label{eq_m2} \ .
\end{eqnarray}
The range of parameters in the various optimizations are such that
$ 2 \ell -1 \le x+y+z \le \ell$. The step between Eqs.~(\ref{eq_m1}) 
and (\ref{eq_m2}) amounts to maximize $m$ over $z$, which can be done 
analytically. It
is then very easy to determine the function $\widehat{m}(\ell,y)$ numerically.
Finally, defining $S(\ell)=\tau(\ell)-2\sigma(\ell)$, we determined
numerically this function (see Fig.~\ref{fig_2nd_moment})
and found that $S>0$ for all parameters such that
$\sigma>0$: the second moment of ${\cal N}_L(G)$
is then exponentially larger than the square of the first moment,
which forbids the use of the second moment method to 
determine the typical value of ${\cal N}_L$.

\subsubsection{Arbitrary connectivity distribution and regular}
The computation of ${\cal P}_{LXYZ}$ in the configuration model can be done
similarly to the one of ${\cal P}_L$ (cf. Eq.~(\ref{eq_combin_PL_arb})). To 
simplify notations let us define $U=2L-3X-Y-2Z$, $V=2X$, and the multinomial
coefficient
\begin{equation}
{N \choose U,V,Z} = \frac{N!}{U! V! Z! (N-U-V-Z)!} \ ,
\end{equation}
for $U+V+Z \le N$. We also use $(k)_n=k(k-1)\dots(k-n+1)$.
With these conventions one finds
\begin{equation}
{\cal P}_{LXYZ} = \frac{1}{{N \choose U,V,Z}} \left( 
\underset{\{U_k, V_k, Z_k \}}{\sum}
\prod_{k=2}^\infty {N q_k\choose U_k,V_k,Z_k} (k)_2^{U_k} 
(k)_3^{V_k} (k)_4^{Z_k}  \right) 
\frac{(cN-2(2L-Y)-1)!!}{(cN-1)!!} \ .
\end{equation}
Indeed, $U$ (resp. $V$, $Z$) is the number of vertices with two 
(resp. three, four) half-edges involved in the pattern, and $(U_k,V_k,Z_k)$ the
number of such vertices among the ones of degree $k$. In consequence the sum
is over non-negative integers with $V_2=Z_2=Z_3=0$, $U_k+V_k+Z_k \le N_k$,
and $\sum_k U_k = U$, $\sum_k V_k = V$, $\sum_k Z_k = Z$. 
These last three constraints can be implemented using the complex integral
representation of Kronecker's delta, themselves evaluated by the saddle point
method in the thermodynamic limit:
\begin{eqnarray}
{\cal P}_{LXYZ} &\doteq&  N^{Y-2L} \exp [ N p(\ell,x,y,z) ]  \ , \\
p(\ell,x,y,z) &=& \frac{1}{2} h(c-4 \ell + 2 y) - \frac{1}{2} h(c) + 2 \ell -y
+ h(2 \ell - 3x -y -2 z) + h(2x) +h(z) + h(1-2\ell+x+y+z) \nonumber \\
&+&  \underset{\theta_1,\theta_2,\theta_3}{{\rm ext}} \left[ \sum_{k=2}^\infty
q_k \ln(1 + (k)_2 \theta_1 + (k)_3 \theta_2 + (k)_4 \theta_3 )
- (2\ell -3x - y -2x)\ln \theta_1 - 2x \ln \theta_2 - z \ln \theta_3 \right]
\ . \nonumber
\end{eqnarray}
Once this function has been determined for a given degree distribution,
the exponential order of $\overline{{\cal N}_L^{*2} (G)}$ can be computed as
\begin{equation}
\overline{{\cal N}_L^{*2} (G)} \doteq \exp [ N \tau(\ell) ] \ , \quad
\tau(\ell) = \underset{x,y,z}{\max} [ p(\ell,x,y,z) + m(\ell,x,y,z)] \ ,
\end{equation}
where $m$ is given in Eq.~(\ref{eq_combin_m2}).

In the regular case, the maximization over the 6 parameters can be performed
analytically, and yields $\tau(\ell) = 2 \sigma(\ell)$~\cite{Garmo}, proving
the concentration (at the exponential order) of ${\cal N}_L(G)$ around its
mean. We expect that for any (fastly decaying) connectivity distribution not 
strictly concentrated on a single integer, $\tau(\ell) > 2 \sigma(\ell)$
when $\sigma(\ell)>0$. A proof of this conjecture would be a quite painful
exercise in analysis that we did not undertake. We however verified numerically
this statement for the Hamiltonian circuits of random graphs with an equal
mixture of vertices of degree 3 and 4, yielding 
$\tau(1) - 2 \sigma(1) \approx 0.002$. 

We have been rather loose in treating the algebraic prefactor hidden
in $\doteq$ for the various expressions of $\overline{{\cal N}_L^2(G)}$.
However it is rather simple to determine the power of $N$ in this prefactor,
collecting the contributions which arise from the Stirling expansions, the
transformation of sums into integrals, and the evaluation of the latter with
the saddle point method. This leads to
\begin{equation}
\frac{\overline{{\cal N}_L^2(G)}}{\overline{{\cal N}_L(G)}^2} = {\rm cst}
\ (1+O(N^{-1})) \exp [N(\tau(\ell)-2\sigma(\ell))] \ ,
\end{equation}
as we observed numerically in Sec.~\ref{sec_enumerations}.

Note also that some informations on the structure of the space of 
configurations can be obtained from this kind of computations. The average
number of pairs of circuits at a given ``overlap'' (number of common edges)
is indeed obtained from the second moment computations if the parameter $y$
is kept fixed.

\subsection{On the union of vertex disjoint circuits}
\label{app_disjoint}
In the statistical mechanics treatment of the main part of the text we used
a model which counts the number ${\cal N}'_L(G)$ of subgraphs of $G$ made of
the union of vertex disjoint circuits of total length $L$. We want to show 
in this appendix that, at the leading exponential order, the average of
${\cal N}'_L(G)$ equals the one of ${\cal N}_L(G)$ in the various ensembles
considered in this appendix. Let us denote ${\cal C}'_L$ the set of subgraphs
of the complete graph on $N$ vertices made of unions of vertex disjoint 
circuits of total length $L$, and ${\cal M}'_L$ its cardinality. As such
subgraphs are still made of $L$ edges connecting $L$ vertices, 
$\overline{{\cal N}'_L(G)} = {\cal M}'_L {\cal P}_L$, where the probability
${\cal P}_L$ is the one defined previously for the computation of 
${\cal N}_L(G)$. Let us define ${\cal M}'_{L,A}$ the cardinality of the
subset of ${\cal C}'_L$ where the subgraphs are made of $A$ disjoint circuits.
A short reasoning leads to
\begin{equation}
{\cal M}'_{L,A} = \frac{N!}{(N-L)!} \frac{1}{2^A} 
\underset{A_3,\dots,A_L}{\sum} \frac{1}{\prod_{i=3}^L A_i! i^{A_i}} 
\ \delta_{\sum_i A_i,A} \ \delta_{\sum_i i A_i,L} \ ,
\end{equation}
where the integers $A_i$ are the number of circuits of length $i$ in the
subgraph. From this expression it is easy to check that 
${\cal M}'_{L,1}={\cal M}_L$, and that as long as $A$ is finite in the 
thermodynamic limit, ${\cal M}'_{L,A} \doteq {\cal M}_L$. More precisely,
one can show that the leading behaviour of ${\cal M}'_L=\sum_A {\cal M}'_{L,A}$
is not modified by contributions with $A$ growing with $N$. Indeed,
\begin{equation}
\frac{{\cal M}'_L}{{\cal M}_L} = 2L \ [t^L]\  \exp\left[ \frac{1}{2}\left(
\frac{t^3}{3} + \dots \frac{t^L}{L}
\right) \right] \ ,
\end{equation}
where $[t^x] f(t)$ denotes the coefficient of order $x$ in the series 
expansion of $f(t)$. Evaluating the right hand side with the saddle point 
method when $L \to \infty$, one can conclude that 
${\cal M}'_L \doteq {\cal M}_L$ and from the above remark 
$\overline{{\cal N}'_L(G)} \doteq \overline{{\cal N}_L(G)}$. As far as annealed
computations are concerned, the distinction between circuits of (extensive)
length $L$ and union of disjoint circuits of total length $L$ does not modify
the entropy. The hypothesis made in the main part of the text is that this
remains true for the quenched computations.

\section{Analysis of the leaf removal algorithm on random graphs with 
arbitrary connectivity distribution}
\label{sec_app_leaf}

We want to justify in this appendix the geometric interpretation of the
null messages elimination we gave in Sec.~\ref{sec_quenched}.
Consider a random graph drawn uniformly among the ones with the
connectivity distribution $q_k$. The 2-core of a graph is the largest
subgraph in which all vertices have connectivity at least two. It can be
determined using the following leaf removal algorithm, which reduces
iteratively the graph. At each time step, if there is at least one
vertex of degree 1, choose randomly one of them, and remove the unique edge
to which it belongs. When there is no vertex of degree 1, the algorithm stops.
At this point either all the edges have been removed and one is left with $N$ 
isolated vertices, or there remains some isolated vertices and a subgraph in 
which all vertices have at least degree 2, i.e. the 2-core of the
initial graph. 

One can define more generally the $q$-core of a graph as the largest subgraph
with minimal degree $q$. For Erd\"os-R\'enyi random graphs, the thresholds
for the appearance of giant $q$-cores have been obtained in~\cite{PiSpWo}.
These results have been recently extended to random graphs with arbitrary
connectivity distributions in~\cite{FeRa}, this appendix can thus be viewed
as an informal presentation of these mathematical works, with the emphasis
put on the quantitative results instead of the mathematical rigor (see also
\cite{kcore_Doro,kcore_Doro2} for an heuristic derivation in the arbitrary
connectivity distributions case, and \cite{kcore_Riordan,kcore_Janson}
for new mathematical treatments of the problem).
In the following we shall study the behaviour of the leaf removal 
algorithm through differential equations for the evolution of the average
connectivity distribution along the execution of the leaf removal. This
method is widely used in mathematics and computer science, see in particular
\cite{Wormald_diff} for a general presentation, and a detailed derivation
of the equations (\ref{eq_dotr}).

We shall denote $T$ the number of steps (elimination of one edge) already 
performed by the algorithm, and $t=T/N$ the reduced time variable. Let
us call $R_k(t) = N r_k(t)$ the average (over the choice of the initial graph 
and the random decisions taken by the algorithm) number of sites of 
connectivity $k$ in the residual graph obtained after $Nt$ time steps of
the algorithm. The initial condition reads obviously $r_k(t=0)=q_k$. If one
calls $\tilde{r}_k(t)$ the probability that the neighbor of the selected 
degree 1 vertex has connectivity $k+1$, the average evolution of the $R$'s 
during the time-step $t \to t+1/N$ reads
\begin{equation}
R_k(t+ 1/N) - R_k(t) = \delta_{k,0} - \delta_{k,1}
+ \sum_{k'=0}^\infty \tilde{r}_{k'}(t) [-\delta_{k,k'+1} + \delta_{k,k'}] \ .
\label{eq_Rk}
\end{equation}
To close this set of equations we have to express the offspring probabilities 
$\tilde{r}_k(t)$ in terms of the connectivity distribution $r_k(t)$. 
As the graph 
is sequentially exposed by the algorithm, the residual graph at time $t$ is
still uniformly distributed according to the connectivity distribution 
$r_k(t)$, hence
\begin{equation}
\tilde{r}_k(t) = \frac{(k+1)r_{k+1}(t)}{\sum_k k r_k(t)} = 
\frac{(k+1)r_{k+1}(t)}{c-2t} \ .
\end{equation}
In the last equality $c$ is the initial mean connectivity $\sum_k k q_k$,
which is reduced by $2/N$ at each time-step. In the thermodynamic limit 
the discrete time relations (\ref{eq_Rk}) become ordinary differential 
equations,
\begin{eqnarray}
\dot{r}_0(t) &=& 1 + \frac{r_1(t)}{c \eta(t)^2} \ , \nonumber \\
\dot{r}_1(t) &=& -1 - \frac{r_1(t)}{c \eta(t)^2} 
+ \frac{2 r_2(t)}{c \eta(t)^2} \ , \nonumber \\
\dot{r}_k(t) &=& - \frac{k r_k(t)}{c \eta(t)^2} 
+ \frac{(k+1) r_{k+1}(t)}{c \eta(t)^2} \quad {\rm for} \ k\ge 2 \ .
\label{eq_dotr}
\end{eqnarray}
where dotted quantities are derivatives with respects to time, and we introduced
the notation $\eta(t) = \sqrt{1 - \frac{2t}{c}}$.

For simplicity let us first assume the existence of a cutoff $k_{\rm m}$ in the
original distribution $q_k$, $q_k = 0$ for $k> k_{\rm m}$. As the leaf
removal procedure never increases the connectivity of one site, this
cutoff remains present in $r_k(t)$ for all times. The equations of rank 
$k_{\rm m}$ in the hierarchy (\ref{eq_dotr}) is then closed on $r_{k_{\rm m}}$.
Using the fact that $\dot{\eta}(t) = -1/(c \eta(t))$, it can be written as
\begin{equation}
\dot{r}_{k_{\rm m}}(t) = -k_{\rm m} \frac{\dot{\eta}(t)}{\eta(t)} r
_{k_{\rm m}}(t) \ ,
\end{equation}
and easily integrated with the initial condition 
$r_{k_{\rm m}}(t=0)=q_{k_{\rm m}}$ as
\begin{equation}
r_{k_{\rm m}}(t) = q_{k_{\rm m}} \eta(t)^{k_{\rm m}} \ .
\end{equation}
Now one can prove by a decreasing recurrence on $k$ from $k_{\rm m}$ down to 2
that
\begin{equation}
r_k(t) = \sum_{n=k}^{k_m} q_n {n \choose k} \eta(t)^k (1-\eta(t))^{n-k}  
\label{eq_solr}
\end{equation}
solves the hierarchy of equations (\ref{eq_dotr}). 
Note also that the 
initial conditions $r_k(0)=q_k$ are enforced by Eq.~(\ref{eq_solr}) as 
$\eta(0)=1$. Once $r_k(t)$ has
been computed for $k\ge 2$, the equation on $r_1$ yields
\begin{equation}
r_1(t) = - c \, \eta(t) (1-\eta(t)) + \sum_{n=1}^{k_m} n \, q_n \, 
\eta(t) (1-\eta(t))^{n-1} \ .
\end{equation}
Finally $r_0(t)$ can be obtained from the normalization condition of the 
$r_k$'s. 

We introduced the cutoff $k_{\rm m}$ to have an explicit starting point of
the downwards recurrence on $k$. However, the expression (\ref{eq_solr})
formally solves the hierarchy of equations (\ref{eq_dotr}) even for
unbounded distributions $q_k$, we shall therefore send the cutoff to infinity
from now on, assuming that all the sums remain convergent.

The 2-core is found when the leaf-removal algorithm stops, at the smallest 
time $t_*$ for which the number of degree 1 vertices vanishes, $r_1(t_*)=0$. 
This equation always admits $t_*=c/2$ as a solution (the graph has then been 
emptied), however if there is a smaller solution the 2-core is non
trivial and the algorithm stops before having removed all edges. 
Calling $\eta=\eta(t_*)$, one obtains if $\eta>0$ :
\begin{equation}
1- \eta = \sum_{k=1}^\infty \frac{k q_k}{c} (1 - \eta)^{k-1} 
= \sum_{k=0}^\infty \tilde{q}_k (1-\eta)^k \ ,
\end{equation}
which is nothing but Eq.~(\ref{eq_eta}) on the fraction of non vanishing
messages obtained in the main part of the text. The confirmation of the
interpretation given in Sec.~\ref{sec_quenched} follows easily:
\begin{itemize}
\item[$\bullet$]
the number of edges in the 2-core is equal to the initial number of edges minus
the number of steps performed by the algorithm before stopping,
$M_{\rm core} = M-N t_* = M \eta^2$.
\item[$\bullet$]
the distribution of the connectivities of the sites in the 2-core is 
$r_k(t_*)$, as expected from Eq.~(\ref{eq_r}).
\end{itemize}

For completeness we also give the number of sites in the 2-core:
\begin{equation}
N_{\rm core} = N \ell_{\rm core} \ , \qquad 
\ell_{\rm core} = \sum_{k=2}^\infty r_k(t_*) = 
 1- \sum_{k=0}^\infty q_k (1-\eta)^k - c \eta(1-\eta) \ .
\label{eq_lcore}
\end{equation}
As it should, this number is smaller than the size of the giant component,
which reads~\cite{MR,NeStWa}:
\begin{equation}
N_{\rm giant} = N \left( 1- \sum_{k=0}^\infty q_k (1-\eta)^k \right) \ .
\end{equation}
Moreover the fraction of sites which are in the giant component but out of
the 2-core is proportional to $\eta(1-\eta)$. Indeed, the corresponding
edges bear exactly one non null directed message, in the formalism of 
Sec.~\ref{sec_quenched}: if both messages were non null the edge would be in
the 2-core, if both vanished the edge would be out of the giant component.

\section{An alternative derivation of $\ell_{\rm max}=1$ when the
minimal connectivity is 3}
\label{app_lmax}
This appendix presents, as a consistency check, another derivation
of the identity $\ell_{\rm max}=1$ for random graph ensemble with minimal
connectivity of 3. In the main part of the text (Sec.~\ref{sec_binf_without}),
we obtained it by inspection on the behaviour of the free-energy in the
large $u$ limit, because of the absence of hard fields.
There exists however another expression of $\ell_{\rm max}$,
in terms of the distribution of evanescent fields, obtained
by taking the large $u$ limit in Eq.~(\ref{eq_lbeta}):
\begin{equation}
\ell_{\rm max} = \frac{c}{2} \int_0^\infty dx_1 \, V_0(x_1) \, dx_2 \, 
V_0(x_2) \frac{x_1 x_2}{1 + x_1 x_2} \ .
\end{equation}
Let us introduce the following
functional of any probability distribution law $A$,
\begin{equation}
F_k[A](x) = \int_0^\infty dx_1 \, A(x_1) \dots
dx_k \, A(x_k) \, \delta(x-h_k(x_1,\dots,x_k)) \ ,
\end{equation}
such that Eq.~(\ref{eq_Rx}) can be rewritten in a compact way as
$V_0 = \sum_k \tilde{q}_k F_k[V_0]$, and  
a bilinear form on the space of probability distribution
functions,
\begin{equation}
\langle A,B \rangle = \int_0^\infty dx \, dy \, A(x) B(y) \frac{xy}{1+xy} \ .
\end{equation}
Consider now this form with its arguments being a distribution $A$ and its 
image through the functional $F_k$:
\begin{equation}
\langle A, F_k[A] \rangle = \int_0^\infty dx_0 \, A(x_0)\, dx_1 \, A(x_1) \dots
dx_k \, A(x_k)  \frac{x_0 h_k(x_1,\dots,x_k)}{1+x_0 h_k(x_1,\dots,x_k)} \ .
\end{equation}
The rational fraction in the integral can be transformed in the following way:
\begin{equation}
\frac{x_0 h_k(x_1,\dots,x_k)}{1+x_0 h_k(x_1,\dots,x_k)} = 
\frac{x_0 \overset{k}{\underset{i=1}{\sum}} x_i }
{x_0 \overset{k}{\underset{i=1}{\sum}} x_i + 
\underset{1\le i <j \le k}{\sum} x_i x_j} =
\frac{x_0 \overset{k}{\underset{i=1}{\sum}} x_i }
{\underset{0\le i <j \le k}{\sum} x_i x_j} \ .
\end{equation}
Both the denominator of this fraction and the 
integration measure $\prod_{i=0}^{k} dx_i \, A(x_i)$ being invariant under 
the permutations of the $k+1$ $x_i$'s, the integral can be computed
by symmetrizing the numerator of the fraction. The normalization of $A$
then gives
\begin{equation}
\langle A, F_k[A] \rangle = \frac{2}{k+1} \ .
\label{eq_id_ps}
\end{equation}
The proof of $\ell_{\rm max}=1$ is now straightforward: 
\begin{equation}
\ell_{\rm max} = \frac{c}{2} \langle V_0,V_0\rangle
= \sum_{k=2}^\infty \frac{c \tilde{q}_k}{2} \langle V_0, F_k[V_0] \rangle \ .
\end{equation}
Using the identity (\ref{eq_id_ps}) and the relation between $\tilde{q}$ and
$q$ (cf. Eq.~(\ref{eq_qtilde_q})), $\ell_{\rm max}$ is found to be the sum of 
$q_k$ for $k \ge 3$, and hence is equal to 1 by normalization.

We also verified numerically that in presence of degree 2 vertices, and hence 
of non trivial hard fields, the limit of Eq.~(\ref{eq_lbeta}) which involves
both evanescent and hard fields, coincide with the expression 
Eq.~(\ref{eq_lmax}) in terms of hard fields only. We believe this could be 
proved analytically, yet we did not find a simple way to do it.


\begin{thebibliography}{99}

\bibitem{Bollo_book} B.~Bollob\'as, \emph{Random graphs},
Cambridge University Press (2001).

\bibitem{Janson_book} S.~Janson, T.~Luczak and A.~Rucinski, 
\emph{Random graphs}, John Wiley and sons (2000).

\bibitem{Wormald_review}  N.~C.~Wormald,  Models of random regular graphs, in
Surveys in Combinatorics, J.D. Lamb and D.A. Preece, eds, 
London Mathematical Society Lecture Note Series {\bf 276}, 239 (1999). 

\bibitem{Robinson_ham} R.~W.~Robinson and N.~C.~Wormald, 
Random Struct. Alg. {\bf 5}, 363 (1994).

\bibitem{Janson_ham} S.~Janson, Comb. Probab. Comput. {\bf 4} 369 (1995).

\bibitem{Garmo} H.~Garmo, Random Struct. Alg. {\bf 15}, 43 (1999).

\bibitem{Janson_distrib} S.~Janson, Combin. Probab. Comput. {\bf 12}, 
27 (2003).

\bibitem{Luczak} T.~Luczak, Random. Struct. Alg. {\bf 2}, 421 (1991).

\bibitem{Flaj} P.~Flajolet, D.~E.~Knuth and B.~Pittel, Discrete Math.
{\bf 75}, 167 (1989).

\bibitem{Frieze} A.~Frieze, Discrete Math. {\bf 59}, 243 (1986).

\bibitem{Posa} L.~P\'osa, Discrete Math. {\bf 14}, 359 (1976).

\bibitem{Johnson} D.~B.~Johnson, SIAM J. Comput. {\bf 4}, 77 (1975).

\bibitem{GaJo}  M.~R.~Garey and D.~S.~Johnson, Computers and Intractability: 
A Guide to the Theory of NP-Completeness. New York: W. H. Freeman, (1983).

\bibitem{proba_Hamilton} M.~E.~Dyer, A.~Frieze and M.~R.~Jerrum, 
SIAM Journal on Computing {\bf 27}, 1262 (1998).

\bibitem{AlBa} R.~Albert and A.-L.~Barabasi, 
Rev. Mod. Phys. {\bf 74}, 47 (2002).

\bibitem{BiCaCo} G.~Bianconi, G.~Caldarelli and A.~Capocci, 
Phys. Rev. E {\bf 71}, 066116 (2005).
 
\bibitem{ben}
H.D.~Rozenfeld, J.E.~Kirk, E.M.~Bollt and D.~Ben-Avraham,
J. Phys. A {\bf 38}, 4589 (2005).

\bibitem{BenKra}  E.~Ben-Naim and P.~L.~Krapivsky,
Phys. Rev. E {\bf 71}, 026129 (2005).

\bibitem{BiMa} G.~Bianconi and M.~Marsili, J. Stat. Mech. P06005 (2005).

\bibitem{MePa_Bethe} M.~M\'ezard and G.~Parisi, Eur. Phys. J. B 
{\bf 20}, 217 (2001).

\bibitem{EPL} E.~Marinari, R.~Monasson and G.~Semerjian, 
Europhys. Lett. {\bf 73}, 8 (2006).

\bibitem{MC_enumeration} 
K.~Klemm and P.~F.~Stadler, Phys. Rev. E {\bf 73}, 025101(R) (2006).

\bibitem{Yedidia} J.~S.~Yedidia, W.~T.~Freeman and Y.~Weiss, in 
\emph{Advances in Neural Information Processing Systems} {\bf 13}, 689 (2001).

\bibitem{Tati} S.~Tatikonda and M.~Jordan, in Proceedings of UAI-2002, 493 
(2002).

\bibitem{Heskes} T.~Heskes, Neural Computation  {\bf 16}, 2379 (2004). 

\bibitem{DIMES} http://www.netdimes.org/

\bibitem{Achli} D.~Achlioptas and Y.~Peres, 
Journal of the AMS {\bf 17}, 947 (2004).

\bibitem{Beyond} M.~M\'ezard, G.~Parisi and M.A.~Virasoro, Spin-glass theory
and beyond, World Scientific, Singapore (1987).

\bibitem{MR}  M.~Molloy and B.~Reed, Combin. Probab. Comput.
{\bf 7}, 295 (1998).

\bibitem{NeStWa} M.~E.~J.~Newman, S.~H.~Strogatz and D.~J.~Watts, Phys. Rev. E
{\bf 64}, 026118 (2001).

\bibitem{MaMo} E.~Marinari and R.~Monasson, JSTAT P09004 (2004).

\bibitem{polymers} M.~M\"uller, M.~Mezard and A.~Montanari, 
J. Chem. Phys. {\bf 120}, 11233 (2004).

\bibitem{MoPaRi}  A.~Montanari, G.~Parisi and F.~Ricci-Tersenghi,
J. Phys. A {\bf 37}, 2073 (2004).

\bibitem{BM}  O.~Rivoire, G.~Biroli, O.~C.~Martin and M.~M\'ezard,
Eur. Phys. J. B {\bf 37}, 55 (2004).

\bibitem{stab_T0}  T.~Castellani, F.~Krzakala and F.~Ricci-Tersenghi,
Eur. Phys. J. B {\bf 47}, 99 (2005).

\bibitem{extreme} J.P.~Bouchaud and M.~M\'ezard,
J. Phys. A {\bf 30}, 7997 (1997).

\bibitem{Guerra} F.~Guerra, Commun. Math. Phys. {\bf 233}, 1 (2003).

\bibitem{Talagrand} M.~Talagrand, Annals of Mathematics {\bf 163}, 221 (2006).

\bibitem{FrLe} S.~Franz and M.~Leone, J. Stat. Phys. {\bf 111}, 535 (2003).

\bibitem{FrLeTo} S.~Franz, M.~Leone and F.~L.~Toninelli, 
J. Phys. A {\bf 36}, 10967 (2003).

\bibitem{Aldous} D.~Aldous and J.M.~Steele, in Discrete Combinatorial
Probability, H.~Kesten Ed., Springer-Verlag (2003).

\bibitem{Counting} A.~Bandyopadhyay and D.~Gamarnik, {\tt math.PR/0510471}
(2005).

\bibitem{MoRi}  A.~Montanari and T.~Rizzo, J. Stat. Mech. P10011 (2005).

\bibitem{PaSl} G.~Parisi and F.~Slanina, {\tt cond-mat/0512529}.

\bibitem{ChCh} M.~Chertkov and V.Y.~ Chernyak, {\tt cond-mat/0601487} and 
{\tt cond-mat/0603189}.

\bibitem{Olivier} O.~Rivoire, J. Stat. Mech. P07004 (2005).

\bibitem{SP} M.~M\'ezard and R.~Zecchina, Phys. Rev. E {\bf 66}, 056126 (2002).

\bibitem{growing} A.-L.~Barabasi and R.~Albert, Science {\bf 286}, 509 (1999).

\bibitem{BiMa2} G.~Bianconi and M.~Marsili, {\tt cond-mat/0511283} (2005).

\bibitem{matching} R.M.~Karp and M.~Sipser, 
Proceedings of FOCS 1981, 364 (1981).

\bibitem{matching2} J.~Aronson, A.~Frieze and B.~Pittel, 
Random Struct. Alg. {\bf 12}, 111 (1998).

\bibitem{Lenka} L.~Zdeborov\'a and M.~M\'ezard, {\tt cond-mat/0603350}.

\bibitem{Bollobas_k} B.~Bollob\'as, J.H.~Kim and J.~Verstra\"ete,
\emph{Regular subgraphs of random graphs},
Random Struct. Alg., to appear.

\bibitem{Martin} M.~Pretti and M.~Weigt, in preparation.

\bibitem{PiSpWo} B.~Pittel, J.~Spencer and N.~C.~Wormald, 
J. Comb. Theory Ser. B {\bf 67}, 111 (1996).

\bibitem{FeRa} D.~Fernholz and V.~Ramachandran, UTCS technical report TR04-13 
(2004), and preprint (2003), http://www.cs.utexas.edu/$\sim$vlr/pubs.html

\bibitem{kcore_Doro} S.N.~Dorogovtsev, A.V.~Goltsev and J.F.F.~Mendes,
Phys. Rev. Lett. {\bf 96}, 040601 (2006).

\bibitem{kcore_Doro2} A.V.~Goltsev, S.N.~Dorogovtsev and J.F.F.~Mendes,
{\tt cond-mat/0602611} (2006).

\bibitem{kcore_Janson} S.~Janson and M.J.~Luczak, {\tt math.CO/0508453} (2005).

\bibitem{kcore_Riordan} O.~Riordan, {\tt math.CO/0511093} (2005).

\bibitem{Wormald_diff} N.~C.~Wormald, The differential equation method for 
random graph processes and greedy algorithms, in  Lectures on Approximation 
and Randomized Algorithms, M.~Karonski and H.~J.~Proemel, eds, 73 (1999). 

\bibitem{conf_Bollo} B.~Bollob\'as, Europ. J. Combinatorics {\bf 1}, 
311 (1980).

\end{thebibliography}
\end{document}